\documentclass[11pt]{iopart}

\usepackage{iopams}
\usepackage{graphicx}
\usepackage[T1]{fontenc} 

\begin{document} 

\title[Acceleration of a polarized neutron with internal EW forces]{Acceleration of a polarized neutron by internal weak nuclear forces}

\author{M. Donaire}
\address{Departamento de F\'isica Te\'orica, At\'omica y \'Optica,  Universidad de Valladolid, 47011 Valladolid, Spain}
\address{CNRS, Universit\'e Grenoble Alpes, Institut N\'eel, 38000, Grenoble, France}
\ead{manuel.donaire@uva.es}
\begin{abstract}
	It is proven that a polarized neutron gets accelerated by internal nuclear forces along the coherent rotation of its spin. The net force upon the neutron arises from the weak nuclear interactions between its quarks.  It is the result of the simultaneous breaking of parity symmetry by the chiral weak interactions between the neutron’s quarks, and of time-reversal symmetry along the inversion of their spins. The variation of the neutron's kinetic momentum is accompanied with the transfer of an equivalent momentum to the fields of the Z and W-bosons that mediate the interactions, in the opposite direction. The effect is linear in Fermi's constant. Using the simplest hadron models, an upper bound of the order of meters per second is estimated for the velocity variation of the polarized neutron along the spin-flip process. 
\end{abstract}
\begin{small}
\end{small}
\submitto{\jpg}
\maketitle

\section{Introduction}\label{sec:intro}
	
The weak nuclear interaction of the Standard Model is not parity invariant \cite{Donoghe,MIT_bag2_Greiner,Peskin}.  This manifests in the differential coupling of the gauge bosons to left-handed and right-handed fermions. In turn, that results in a net alignment between the spins and the momenta of the hadrons and the leptons which participate in the interaction, giving rise to chiral phenomena. It explains, for instance, the celebrated experiment of Wu \emph{et al.} \cite{Wu} on the beta decay of polarized Co$^{60}$ nuclei,
where left-handed electrons are preferably emitted in the direction opposite to the spins of
the nuclei \cite{betadecay}. 

Along the same lines, in this article we aim to prove that the electroweak (EW) interactions between the quarks of a polarized neutron may generate a net force on the neutron along the rotation of its spin.  On physical grounds this is the result of the alignment between the spins of the quarks and the momenta of the Z and W bosons which mediate their chiral weak interactions. Thus, we will see that the variation of the orientation of the total spin of a neutron, caused for instance by a radio-frequency (RF) magnetic field, is accompanied by a variation of its kinetic momentum. The force responsible of this variation is thus mediated by virtual Z and W-bosons, whose fields carry off an equivalent momentum in the opposite direction, in accord with total momentum conservation.  We will show that this force is linear in $G_{F}$,  causing a variation on the neutron velocity of the order of meters per second along the spin reversal process. In brief, this phenomenon is induced by the same symmetry considerations that explain the directionality of beta emission. Hence an analogy can be drawn between both phenomena. In the beta decay, a net average kinetic momentum is gained by the polarized nuclei as a result of its recoil from the preferential emission of electrons and antineutrinos along the nuclear spin direction. In our case, the force on the polarized neutron is caused by the recoil of its quarks from the variation of the preferential direction of emission of virtual Z and W-bosons along the spin reversal process.


In the reminder of this Introduction we summarize previous findings related to this phenomenon. Homogeneity of space and translation invariance of the vacuum state of any quantum field theory with gauge fields and conserved currents, imply that the net momentum of vacuum field 
fluctuations vanishes identically. However, when field fluctuations couple to current fluctuations through a Hamiltonian which violates both parity and 
time-reversal, it is symmetry allowed for the field fluctuations to carry a net linear momentum.  In virtue of total momentum conservation, 
the currents must gain a kinetic momentum of equal strength in the opposite direction. 
This was first envisaged by A. Feigel in a magnetoelectric medium \cite{Feigel,Croze}. Using a semiclassical approach he showed that, in crossed electric and magnetic fields, 
$\mathbf{E}\perp\mathbf{B}$, the medium generates a magnetoelectric birefringence along $\mathbf{E}\times\mathbf{B}$. As a result, the momenta carried by the modes of
the quantum electromagnetic (EM) field fluctuations propagating parallel and antiparallel to $\mathbf{E}\times\mathbf{B}$ do not cancel each other and, in virtue of total momentum conservation, a momentum equivalent to their difference is transferred to the magnetoelectric medium in the opposite direction. Later, B. van-Tiggelen and collaborators, using a fully 
QED approach, have computed the momentum gained by a hydrogen atom in crossed electric and magnetic fields \cite{Kawka}, and the momentum of a chiral molecule in 
a magnetic field \cite{PRLDonaire}. In the latter case, parity is broken by the chiral 
distribution of the ligands around the active ion of the molecule and time-reversal is broken by the magnetic field. It was found in Ref.\cite{JPCMDonaire} 
that, in its major part, the acceleration of the molecule is accompanied by a net EM momentum associated to the electrostatic and magnetic fields sourced by the charges and the helical currents within the chiral molecule. In either case, the kinetic momenta 
gained by the hydrogen atom and the chiral molecule were found to be too small to be experimentally observed. 

Further, it has been proved in Ref.\cite{JHEP} that an analogous phenomenon may take place in nuclear physics,  where the EW    
interaction is not parity invariant. Using the Fermi four-fermion interaction, it was shown that an unpolarized proton 
in a magnetic field accelerates along that field, transferring an equivalent momentum to virtual positrons and neutrinos. 
The velocity of the proton computed that way happens to be of order $G_{F}^{2}$, thus too small to be detected too. However, the reason for that small value was the inappropriate use of the Fermi interaction in a process that, in contrast to the beta decay, does not involve actual leptons. Hence, the estimate in Ref.\cite{JHEP} is not the leading order one but a second order correction. 

In this article we proceed along the lines of Ref.\cite{JHEP}, but we use the gauge theory of EW interactions instead to calculate the velocity variation undergone by a polarized neutron along its spin reversal.  Using two simple models of hadron confinement, namely, the MIT-bag model 
and the harmonic oscillator model, 
we estimate a velocity variation of the order of meters per second. If confirmed experimentally, this finding would prove the existence of a net internal force and the 
transfer of momentum to virtual particles in chiral systems, and would open up the possibility of probing the hadron structure with low-energy techniques.

\section{Fundamentals}\label{sec:fundam}	

In this Section we describe the Lagrangian and the Hamiltonian of the system, we model the wave function of the polarized neutron and describe the spin-flip procedure driven by a radio-frequency field. 

\subsection{Lagrangian and Hamiltonian}	
For simplicity and without much loss of generality, let us consider the Lagrangian of a neutron as the sum of the Lagrangians of its three valence or constituent quarks, two down and one up, which are bound by some effective QCD confinement potential, $\mathcal{V}_{QCD}$, and are minimally coupled to the EW bosons,
\begin{equation}
\fl	L=\sum_{j=1}^{3}\int d^{3}r_{j}\Bigl[\bar{\Psi}_{j}\bigl(i\gamma^{\mu}D^{j}_{\mu}-m_{j}\mathbb{I}\bigr)\Psi_{j}-\sum_{k<j}^{3}\int d^{3}r_{k}\Psi^{\dagger}_{j}\Psi_{k}\mathcal{V}_{QCD}(\mathbf{r}_{j}-\mathbf{r}_{k})\Psi^{\dagger}_{k}\Psi_{j}\Bigr].\label{Lagrange}
\end{equation}
In this equation, given in natural units, $\Psi$ denotes a generic isospin doublet, $\Psi_{j}=\bigl(\psi^{u}_{j}(t,\mathbf{r}_{j})$ $\psi^{d}_{j}(t,\mathbf{r}_{j})\bigr)$,  and $D^{j}_{\mu}$ is the covariant derivative operator upon quark $j$. In terms of the EW gauge fields it reads,
\begin{eqnarray}
	D^{j}_{\mu}&=\partial_{\mu}^{j}\mathbb{I}_{\textrm{iso}}-ie\:A_{\mu}(\mathbf{r}_{j})\mathbb{Q}-\frac{igZ_{\mu}(\mathbf{r}_{j})}{2\cos{\theta_{W}}}\bigl[\mathbb{I}\bigl(\tau_{3}-2\sin^{2}{\theta_{W}}\mathbb{Q}\bigr)-\gamma_{5}\tau_{3}\bigr]\nonumber\\
	&-\frac{ig}{\sqrt{2}}\bigl[W^{+}_{\mu}(\mathbf{r}_{j})\frac{\mathbb{I}-\gamma_{5}}{2}
	(\tau_{1}+i\tau_{2})+\textrm{h.c.}\bigr],\label{CovarD}
\end{eqnarray}
where $\mathbb{I}_{\textrm{iso}}$, $\mathbb{Q}$ and $\tau_{i}$ are two-by-two matrices acting on the isospin indices, namely, the identity, the electric charge operator and the isospin generators, respectively; $\mathbb{I}$ and $\gamma_{5}$ act on the four-spinor components of $\psi_{j}^{u,d}$; and $e$, $g/(2\cos{\theta_{W}})$, $g/\sqrt{2}$ are the coupling constants of A, Z and W gauge fields, respectively, with $\theta_{W}$ the weak mixing angle.

As for the quark confinement, without much loss of generality and for numerical estimations we will consider two simple models. These are, the MIT-bag model for valence and ultrarelativistic quarks \cite{MIT_bag1}, and the harmonic-oscillator model for constituent and nonrelativistic quarks \cite{Isgur} --see \ref{app1}. In the MIT-bag the three quarks are massless and are confined within a static spherical cavity of radius $R_{n}$ by means of boundary conditions at the surface. These conditions prevent the current flow through the surface and impose the balance between the outwards pressure of the quarks and the inward QCD vacuum pressure. In the nonrelativistic harmonic oscillator, all the quarks have an equivalent mass $m$ and interact pairwise through harmonic potentials of equivalent natural frequencies.

In the remainder of this article we will work within the Hamiltonian formalism in Schr\"{o}dinger's picture. First we identify the non-perturbative Hamiltonians with those of the neutron quarks and the free boson fields. As for the former, it can be derived from Eq.(\ref{Lagrange}),
\begin{eqnarray}
	H_{n}&=\sum_{j=1}^{3}\int d^{3}r_{j}\Bigl[\sum_{f=u,d}\bar{\psi}^{f}_{j}\bigl(-i\gamma^{l}\nabla^{j}_{l}+m_{j}\bigr)\psi^{f}_{j}+\sum_{k<j}^{3}\int d^{3}r_{k}\nonumber\\
	&\times\sum_{\tilde{f}=u,d}\psi^{f\dagger}_{j}\psi^{\tilde{f}}_{k}\mathcal{V}_{QCD}(\mathbf{r}_{j}-\mathbf{r}_{k})\psi^{\tilde{f}\dagger}_{k}\psi^{f}_{j}\Bigr].\label{Ho}
\end{eqnarray}
As for the free EW bosons,  the gauge fields $A_{\mu}$, $Z_{\mu}$ and $W^{\pm}_{\mu}$, that we will denote generically by $V_{\mu}$, can be decomposed in sums over normal modes of wave vector $\mathbf{q}$,  polarization $\lambda$, creation/annihilation operators $a_{V,\mathbf{q}}^{\lambda(,\dagger)}$, and energy $\mathcal{E}^{V}_{\mathbf{q}}=\sqrt{q^{2}+M^{2}_{V}}$, with $M_{V}$ being the boson mass \cite{Peskin}. Hence, the Hamiltonian of the free bosons reads 
\begin{equation}
	H_{F}^{V}=\sum_{\mathbf{q},\lambda}\mathcal{E}^{V}_{\mathbf{q}}a_{V,\mathbf{q}}^{\lambda,\dagger}a_{V,\mathbf{q}}^{\lambda},\quad V=\{A,Z,W^{\pm}\}.\label{HF}
\end{equation}

The Hamiltonian densities of the interactions between quarks and EW bosons can be read off from the minimal coupling term of Eq.(\ref{Lagrange}) together with Eq.(\ref{CovarD}), 
\begin{eqnarray}
	\mathcal{W}_{EM}=e\sum_{i=1}^{3}j_{EM}^{i,\mu}A_{\mu}(\mathbf{r}_{i}),\qquad\mathcal{W}_{nc}=\frac{g}{2\cos{\theta_{W}}}\sum_{i=1}^{3}j_{nc}^{i,\mu}Z_{\mu}(\mathbf{r}_{i}),\nonumber\\
	\mathcal{W}_{cc}=\frac{g}{\sqrt{2}}\sum_{i=1}^{3}j_{cc}^{i,\mu}W^{+}_{\mu}(\mathbf{r}_{i})+\textrm{h.c.}\label{Ws}
\end{eqnarray}
In these equations $j_{EM}^{i,\mu}$ is the parity-invariant EM current density of quark $i$, $j_{EM}^{i,\mu}=q_{u}\bar{\psi}^{u}_{i}\gamma^{\mu}\psi^{u}_{i}+q_{d}\bar{\psi}_{i}^{d}\gamma^{\mu}\psi^{d}_{i}$, where  $q_{u,d}$ are the electric charges;   
$j_{nc}^{i,\mu}=\bar{\psi}^{d}_{i}\gamma^{\mu}(g_{v}^{d}\mathbb{I}-g_{a}^{d}\gamma_{5})\psi^{d}_{i}+\bar{\psi}^{u}_{i}\gamma^{\mu}(g_{v}^{u}\mathbb{I}-g_{a}^{u}\gamma_{5})\psi^{u}_{i}$ is the weak neutral and chiral current, with $g_{v,a}^{u,d}$ being the vector $(v)$ and axial $(a)$ couplings; and 
$j_{cc}^{i,\mu}=\frac{1}{2}\bar{\psi}^{d}_{i}\gamma^{\mu}(\mathbb{I}-\gamma_{5})\psi^{u}_{i}$  is the weak charged and chiral current. Hereafter we will refer to $j_{nc,cc}^{i,0}$ as chiral charge densities \footnote{This terminology is intended to avoid any possible confusion with the ordinary definition of weak charge.}.\\

Next, we model the wavefunction of a polarized neutron and describe the spin-flip procedure.  We recall that our aim is to prove that, as the neutron spin is inverted in a coherent manner, a variation in the total kinetic momentum of the neutron is caused by the weak nuclear forces between its quarks.


\subsection{Neutron wavefunction}\label{sec:calcul}

For the sake of clarity, we take the simplest three-quark wavefunction of a polarized neutron. We consider that the neutron spin results entirely from the addition of the spins of the wavefunctions of three valence or constituent quarks, upon which we impose the  symmetry conditions. Since the wavefunction must be totally antisymmetric and, in the ground state,  the spatial wavefunction of each quark is an $l=0$ function, the color function must be an antisymmetric SU(3) singlet, $\xi_{color}$, and the spin-flavour wavefunction must be symmetrized ($\mathcal{S}$). All together, the ground state of a spin-down neutron can be written as \cite{Donoghe}
\begin{eqnarray*}
\fl |n^{0}_{\downarrow}\rangle=\frac{\xi_{color}}{\sqrt{3}}\left[|d\downarrow d\uparrow u\downarrow\rangle^{0}_{\mathcal{S}}-2 |d\downarrow d\downarrow u\uparrow\rangle^{0}_{\mathcal{S}}\right]=
\frac{\xi_{color}}{\sqrt{18}}\Bigl[|u\downarrow d\uparrow d\downarrow\rangle^{0}+ |d\downarrow d\uparrow u\downarrow\rangle^{0}
\nonumber\\ 
+ |d\uparrow d\downarrow u\downarrow\rangle^{0}+ |d\uparrow u\downarrow d\downarrow\rangle^{0}+|u\downarrow d\downarrow d\uparrow\rangle^{0} + |d\downarrow u\downarrow d\uparrow\rangle^{0} 
 \nonumber\\
-2\Bigl( |u\uparrow d\downarrow d\downarrow\rangle^{0} + |d\downarrow u\uparrow d\downarrow\rangle^{0}+|d\downarrow d\downarrow u\uparrow\rangle^{0}\Bigr)\Bigr],\label{spinflav}
\end{eqnarray*}
and likewise for $|n^{0}_{\uparrow}\rangle$ with all the spins reversed --the superscript $0$ in $|n^{0}_{\uparrow,\downarrow}\rangle$ refers to the $l=0$ ground state of the spatial wavefunctions. As for the corresponding spatial wavefunction, disregarding the irrelevant color function, it reads
\begin{eqnarray}
\fl \Psi^{0}_{n\downarrow}(\mathbf{r}_{1},\mathbf{r}_{2},\mathbf{r}_{3})=\frac{1}{\sqrt{18}}\Bigl[\psi^{u,-}_{0}(\mathbf{r}_{1})\psi^{d,+}_{0}(\mathbf{r}_{2}) \psi^{d,-}_{0}(\mathbf{r}_{3}) + \psi^{d,-}_{0}(\mathbf{r}_{1})\psi^{d,+}_{0}(\mathbf{r}_{2}) \psi^{u,-}_{0}(\mathbf{r}_{3})\nonumber\\
 + \psi^{d,+}_{0}(\mathbf{r}_{1})\psi^{d,-}_{0}(\mathbf{r}_{2}) \psi^{u,-}_{0}(\mathbf{r}_{3})+\psi^{d,+}_{0}(\mathbf{r}_{1})\psi^{u,-}_{0}(\mathbf{r}_{2}) \psi^{d,-}_{0}(\mathbf{r}_{3})\nonumber\\
+\psi^{u,-}_{0}(\mathbf{r}_{1})\psi^{d,-}_{0}(\mathbf{r}_{2}) \psi^{d,+}_{0}(\mathbf{r}_{3})+\psi^{d,-}_{0}(\mathbf{r}_{1})\psi^{u,-}_{0}(\mathbf{r}_{2}) \psi^{d,+}_{0}(\mathbf{r}_{3})\nonumber\\
-2\Bigl(\psi^{u,+}_{0}(\mathbf{r}_{1})\psi^{d,-}_{0}(\mathbf{r}_{2}) \psi^{d,-}_{0}(\mathbf{r}_{3})+\psi^{d,-}_{0}(\mathbf{r}_{1})\psi^{u,+}_{0}(\mathbf{r}_{2}) \psi^{d,-}_{0}(\mathbf{r}_{3})\nonumber\\
+\psi^{d,-}_{0}(\mathbf{r}_{1})\psi^{d,-}_{0}(\mathbf{r}_{2}) \psi^{u,+}_{0}(\mathbf{r}_{3})\Bigr)\Bigr],\label{Psinup}
\end{eqnarray}
where the subscript $0$ in the wavefunction of each quark denotes its ground state and the superscripts $+$ and $-$ stand for spin-up and spin-down states.
Also, we will find that the intermediate states in the quantum calculation involves the wave function of polarized protons. For a spin-up proton, the expression for $\Psi^{0}_{p\uparrow}$ is the same as that in Eq.(\ref{Psinup})  but for the replacements $d\leftrightarrow u$, $+\leftrightarrow -$ in the quark wavefunctions.

\subsection{Spin-flip procedure}

In order to reverse the spin of the neutron in a coherent manner we employ the $\pi$-pulse of a radio-frequency (RF) 
field. 

Let us consider the neutron initially polarized along the $-\hat{\mathbf{z}}$ direction in the presence of an intense and constant magnetic field 
$\mathbf{B}_{0}=B_{0}\hat{\mathbf{z}}$. 
To invert the spin, an additional RF magnetic field $\mathbf{B}_{1}$ is introduced which rotates at the Larmor frequency $\omega_{0}=-\gamma_{n}B_{0}$ in the $xy$ plane, 
with $\gamma_{n}$ being the neutron gyromagnetic ratio, $\mathbf{B}_{1}(t)=B_{1}[\cos{(\omega_{0}t)}\hat{\mathbf{x}}+\sin{(\omega_{0}t)}\hat{\mathbf{y}}]$. This field is circularly polarized around $\hat{\mathbf{z}}$, and its strength $B_{1}$ is much weaker than 
$B_{0}$.  The resultant external EM vector potential can be written as 
\begin{equation}
\fl\mathbf{A}_{\textrm{ext}}=\frac{B_{0}}{2}(-y\hat{\mathbf{x}}+x\hat{\mathbf{y}})
+\frac{B_{1}}{\omega_{0}}[\cos{(k_{0}z-\omega_{0}t)}\hat{\mathbf{x}}-\sin{(k_{0}z-\omega_{0}t)}\hat{\mathbf{y}}],\quad k_{0}z\rightarrow0,
\end{equation}
with photon density $\mathcal{N}_{\gamma}=B_{1}^{2}/\omega_{0}$. The coupling of these photons to the constituent/valence quarks is through the Hamiltonian 
$\mathcal{W}_{EM}$ in Eq.(\ref{Ws}). All in all that interaction 
simplifies to $W_{EM}\approx-\gamma_{n}\boldsymbol{\sigma}\cdot\mathbf{B}/2$, where $\boldsymbol{\sigma}$ applies to the two-component spin-state of the neutron with basis vectors $\{(|n\uparrow\rangle,0),(0,|n\downarrow\rangle)\}$. Further, replacing $\mathbf{B}$ in $W_{EM}$ with the classical expression of 
$\mathbf{B}_{0}+\mathbf{B}_{1}$ or using the quantum expression for $\mathbf{B}$ and the EM states containing the photons of the field $\mathbf{B}_{1}$ otherwise, one can apply time-dependent perturbation theory and compute an effective Rabi (R) Hamiltonian for the interaction between the RF  field and the neutron. That is,  
\begin{equation}
W^{\textrm{R}}_{EM}\approx -\frac{\omega_{0}}{2}|n\downarrow\rangle\langle n\downarrow| +\frac{\omega_{0}}{2}|n\uparrow\rangle\langle n\uparrow|+ -\frac{\omega_{1}}{2}(|n\downarrow\rangle\langle n\uparrow|+|n\uparrow\rangle\langle n\downarrow|),
\end{equation}
where $\omega_{1}=-\gamma_{n}B_{1}$ is the Rabi frequency. This Hamiltonian accounts for an infinite number of interaction processes involving multiple and alternate absorptions 
and re-emissions of photons of the field $\mathbf{B}_{1}$. Starting with the spin down neutron ground state $|n^{0}_{\downarrow}\rangle$ at the initial time $t=0$, the neutron 
spin wavefunction reads, at a later time $t$,
\begin{equation}
|n^{0}(t)\rangle=\cos{(\omega_{1}t/2)}|n^{0}_{\downarrow}\rangle-i\sin{(\omega_{1}t/2)}e^{-i\omega_{0}t}|n^{0}_{\uparrow}\rangle,\label{nt}
\end{equation}
In turn, this makes the total neutron spin oscillate at frequency $\omega_{1}$, 
\begin{equation}
\langle\mathbf{S}_{n}(t)\rangle=\frac{-1}{2}\left[\sin{(\omega_{1}t)}\sin{(\omega_{0}t)}\hat{\mathbf{x}}-\sin{(\omega_{1}t)}\cos{(\omega_{0}t)}\hat{\mathbf{y}}
+\cos{(\omega_{1}t)}\hat{\mathbf{z}}\right],
\end{equation}
inverting its direction in a time interval $\pi/\omega_{1}$ and absorbing one of the photons of frequency $\omega_{0}$ and momentum $k_{0}\hat{\mathbf{z}}$.

\section{Total momentum conservation}\label{MomentumK}

As argued in the Introduction, the kinetic momentum gained by the neutron along the spin reversal process relies on the conservation of a linear momentum all along the spin-flip process. Thus, the system of hadrons and EW bosons governed by the Hamiltonians of Eqs.(\ref{Ho})-(\ref{Ws}) possesses a conserved linear
momentum  $\mathbf{K}$ which comprises the canonical conjugate momentum of the three quarks bound in the neutron and the \emph{transverse} momenta of the EW bosons, $\mathbf{K}=\mathbf{P}+\mathbf{P}^{A}_{\perp}+\mathbf{P}^{Z}_{\perp}+\mathbf{P}^{W}_{\perp}$.  In this expression the transverse momentum operator of each boson V reads
\begin{equation}
	\mathbf{P}_{\perp}^{V}=\sum_{\mathbf{q},\lambda}\mathbf{q}a_{V,\mathbf{q}}^{\lambda,\dagger}a_{V,\mathbf{q}}^{\lambda},\quad V=\{A,Z,W^{\pm}\},
\end{equation}
whereas the total conjugate momentum $\mathbf{P}$ is made of the total kinetic momentum $\mathbf{P}_{\textrm{kin}}$ and the \emph{longitudinal} momenta associated to the EW bosons, $\mathbf{P}=\mathbf{P}_{\textrm{kin}}+\mathbf{P}^{A}_{\parallel}+\mathbf{P}^{Z}_{\parallel}+\mathbf{P}^{W}_{\parallel}$ \cite{Cohen}. The expressions for the longitudinal momenta in terms of boson fields can be deduced from the spatial components of the  covariant derivative in Eq.(\ref{CovarD}). It suffices to realize that, in the minimal coupling formalism, the total conjugate and kinetic momentum operators read $\mathbf{P}=-i\sum_{j=1}^{3}\boldsymbol{\nabla}^{j}$, $\mathbf{P}_{\textrm{kin}}=-i\sum_{j=1}^{3}\mathbf{D}^{j}$, where $j$ runs over the three quarks, from which, 
\begin{equation}
	\mathbf{P}^{A}_{\parallel}+\mathbf{P}^{Z}_{\parallel}+\mathbf{P}^{W}_{\parallel}=-i\sum_{j=1}^{3}(\boldsymbol{\nabla}^{j}-\mathbf{D}^{j}).
\end{equation}

The proof of the conservation of $\mathbf{K}$ is as follows. Writing the total Hamiltonian $H$ in terms of a Hamiltonian density $\mathcal{H}$ and the Hamiltonian terms of Eqs.(\ref{Ho},\ref{HF},\ref{Ws}), 
\begin{small}
\begin{eqnarray}
\fl	H&=\sum_{j=1}^{3}\int d^{3}r_{j}\sum_{f=u,d}\psi^{f\dagger}_{j}\mathcal{H}\psi^{f}_{j}\nonumber\\
\fl	&=\sum_{j=1}^{3}\int d^{3}r_{j}\Bigl[\sum_{f=u,d}\bar{\psi}^{f}_{j}\bigl(-i\gamma^{l}\nabla^{j}_{l}+m_{j}\bigr)\psi^{f}_{j}+\sum_{k<j}^{3}\int d^{3}r_{k}
	\sum_{\tilde{f}=u,d}\psi^{f\dagger}_{j}\psi^{\tilde{f}}_{k}\mathcal{V}_{QCD}(\mathbf{r}_{j}-\mathbf{r}_{k})\psi^{\tilde{f}\dagger}_{k}\psi^{f}_{j}\Bigr]\label{h1}\\
\fl	&+\sum_{j=1}^{3}\int d^{3}r_{j}\Bigl[e\:j_{EM}^{j,\mu}A_{\mu}(\mathbf{r}_{j})+\frac{g}{2\cos{\theta_{W}}}j_{nc}^{j,\mu}Z_{\mu}(\mathbf{r}_{j})+\frac{g}{\sqrt{2}}\Bigl(j_{cc}^{j,\mu}W^{+}_{\mu}(\mathbf{r}_{j})+\textrm{h.c.}\Bigr)\Bigr]\label{h2}\\
\fl	&+\sum_{V=\{A,Z,W^{\pm}\}}\sum_{\mathbf{q},\lambda}\mathcal{E}^{V}_{\mathbf{q}}a_{V,\mathbf{q}}^{\lambda,\dagger}a_{V,\mathbf{q}}^{\lambda},\label{h3}
\end{eqnarray}
\end{small}
the statement of momentum conservation reads $[\mathbf{K},\mathcal{H}]=\mathbf{0}$. On the other hand, the canonical commutation relations of the total canonical linear momentum $\mathbf{P}$ and  the bosonic creation and annihilation operators of the vector fields with any operator $\mathcal{O}$ read,
\begin{equation}
\fl	[\mathbf{P},\mathcal{O}]=-i\sum_{j=1}^{3}\boldsymbol{\nabla}_{j}\mathcal{O},\quad[a_{V,\mathbf{q}}^{\lambda},\mathcal{O}]=\frac{\partial\mathcal{O}}{\partial a_{V,\mathbf{q}}^{\lambda,\dagger}},\quad[a_{V,\mathbf{q}}^{\lambda,\dagger},\mathcal{O}]=-\frac{\partial\mathcal{O}}{\partial a_{V,\mathbf{q}}^{\lambda}},\quad V=\{A,Z,W^{\pm}\}.
\end{equation}
It is straightforward to verify that the terms of  Eqs. (\ref{h1}) and (\ref{h3}) commute with $\mathbf{P}$ and with the transverse momenta of the gauge fields. As for the terms in Eq.(\ref{h2}), things are more complicated because of the presence of the gauge fields in the interaction. Let us write the gauge fields at the position vector of some valence or constituent quark, $\mathbf{r}_{j}$, as series over normal modes, creation and annihilation operators and polarization vectors of each field,
\begin{equation}
	\mathbf{V}(\mathbf{r}_{j})=\sum_{\mathbf{q},\lambda}Q_{V}\Bigl[\boldsymbol{\epsilon}_{\mathbf{q},\lambda}a_{V,\mathbf{q}}^{\lambda}e^{i\mathbf{q}\cdot\mathbf{r}_{j}}+\boldsymbol{\epsilon}^{*}_{\mathbf{q},\lambda}a_{V,\mathbf{q}}^{\lambda,\dagger}e^{-i\mathbf{q}\cdot\mathbf{r}_{j}}\Bigr],\quad \mathbf{V}=\{\mathbf{A},\mathbf{Z},\mathbf{W}^{\pm}\},
\end{equation}
where $Q_{V}$ is a prefactor specific to each gauge field $V$ and $\boldsymbol{\epsilon}^{(*)}_{\mathbf{q},\lambda}$ are the polarization vectors. Next let us compute the commutators of the total canonical momentum, $\mathbf{P}$, and the transverse momentum of a given gauge field $\mathbf{Y}$, $\mathbf{P}_{\perp}^{Y}$,  with any  gauge field $\mathbf{V}$ at the position of a given quark, $\mathbf{r}_{a}$, $\mathbf{Y},\mathbf{V}\in\{\mathbf{A},\mathbf{Z},\mathbf{W}^{\pm}\}$, $a=1,2,3$,
\begin{small}
\begin{equation}
\fl	[\mathbf{P},\mathbf{V}(\mathbf{r}_{a})]=-i\sum_{j=1}^{3}\boldsymbol{\nabla}_{j}\mathbf{V}(\mathbf{r}_{a})=\sum_{\mathbf{q},\lambda}\mathbf{q}Q_{V}\Bigl[\boldsymbol{\epsilon}_{\mathbf{q},\lambda}a_{V,\mathbf{q}}^{\lambda}e^{i\mathbf{q}\cdot\mathbf{r}_{a}}-\boldsymbol{\epsilon}^{*}_{\mathbf{q},\lambda}a_{V,\mathbf{q}}^{\lambda,\dagger}e^{-i\mathbf{q}\cdot\mathbf{r}_{a}}\Bigr],\\
\end{equation}
\begin{equation}	
\fl	[\mathbf{P}_{\perp}^{Y},\mathbf{V}(\mathbf{r}_{a})]=\sum_{\mathbf{q},\lambda}\Bigl[a_{Y,\mathbf{q}}^{\lambda,\dagger}\frac{\mathbf{V}(\mathbf{r}_{a})}{\partial a_{Y,\mathbf{q}}^{\lambda,\dagger}}-a_{Y,\mathbf{q}}^{\lambda}\frac{\mathbf{V}(\mathbf{r}_{a})}{\partial a_{Y,\mathbf{q}}^{\lambda}}\Bigr]=\sum_{\mathbf{q},\lambda}\mathbf{q}Q_{V}\Bigl[\boldsymbol{\epsilon}^{*}_{\mathbf{q},\lambda}a_{V,\mathbf{q}}^{\lambda,\dagger}e^{-i\mathbf{q}\cdot\mathbf{r}_{a}}-\boldsymbol{\epsilon}_{\mathbf{q},\lambda}a_{V,\mathbf{q}}^{\lambda}e^{i\mathbf{q}\cdot\mathbf{r}_{a}}\Bigr]\delta_{V,Y}.\nonumber
\end{equation}
\end{small}
From the above equations we arrive at $[\mathbf{P}+\mathbf{P}_{\perp}^{V},\mathbf{V}(\mathbf{r}_{a})]=\mathbf{0}\otimes\mathbf{0}$ $\forall$ $\mathbf{V},a$. It follows that the terms of the Hamiltonian in Eq.(\ref{h2}) commute also with $\mathbf{K}$, which completes the proof of $[\mathbf{K},\mathcal{H}]=\mathbf{0}$.

Conservation of total momentum $\mathbf{K}$ implies that the variation $\Delta$ of the total kinetic momentum of the neutron is compensated at any time by an equivalent variation of the momenta of the boson fields with opposite sign, i.e.,
\begin{equation}
	\Delta\langle\mathbf{P}_{\textrm{kin}}\rangle=-\Delta\langle\mathbf{P}^{A}_{\parallel}+\mathbf{P}^{Z}_{\parallel}+\mathbf{P}^{W}_{\parallel}+\mathbf{P}^{A}_{\perp}+\mathbf{P}^{Z}_{\perp}+\mathbf{P}^{W}_{\perp}\rangle.\label{Pconv}
\end{equation}
Accordingly,  associated to the variation of the kinetic momentum there exists an EW force, $\langle\mathbf{F}\rangle$ \cite{force},
\begin{eqnarray}
\fl\qquad\langle \mathbf{F}(t)\rangle&=\frac{d}{dt}\langle\mathbf{P}^{\textrm{kin}}(t)\rangle=-\frac{d}{dt}\left[\langle\mathbf{P}^{\textrm{Z}}_{\parallel}(t)\rangle+\langle \mathbf{P}^{\textrm{W}}_{\parallel}(t)\rangle+\langle\mathbf{P}^{\textrm{Z}}_{\perp}(t)\rangle+\langle\mathbf{P}^{\textrm{W}}_{\perp}(t)\rangle\right]\nonumber\\
\fl&\equiv	\langle\mathbf{F}_{\parallel}(t)\rangle+\langle\mathbf{F}_{\perp}(t)\rangle.\label{primera}
\end{eqnarray}
In the last equality we have identified the force components associated to the time variation of the longitudinal and the transverse momenta of the gauge bosons. The former corresponds to an non-conservative force, while the latter is a conservative force which derives from the gradient of the EW interaction potential \cite{force} --see  \ref{appforce} for an explanation of this identification.

\section{EW momentum calculation}\label{EWmomentum}

In this Section we compute the kinetic momentum gained by the neutron throughout the spin-flip (sf) process, $\Delta\langle\mathbf{P}^{\textrm{sf}}_{\textrm{kin}}\rangle$. To this aim we make use of the conservation of the total momentum $\mathbf{K}$.

In the first place, since the EM interaction is parity invariant, the transfer of EM momentum is only caused by the absorption of an actual RF photon of frequency $\omega_{0}$. That is,  
\begin{equation}
\langle n^{0}_{\downarrow},N_{\gamma}|\mathbf{P}_{\perp}^{A}|n^{0}_{\downarrow},N_{\gamma}\rangle-\langle n^{0}_{\uparrow},N_{\gamma}-1|\mathbf{P}_{\perp}^{A}|n^{0}_{\uparrow},N_{\gamma}-1\rangle=k_{0}\hat{\mathbf{z}}.
\end{equation} 
Typical values for $\omega_{0}$ in experiments range between $10^5$ and $10^8\:$Hz, which corresponds to a variation of the neutron velocity of the order of $10^{-10}$-$10^{-7}\:$m/s, which we will see is negligibly small in comparison to that coming from the virtual Z and W-bosons.

In contrast, the weak nuclear interaction is chiral and its mediating Z and W-bosons are only virtual, so in order for them to carry a net momentum it is the weak interaction between the constituent quarks that should provide them with it. 
We will find that this is indeed the case, and that it is the dominant contribution to Eq.(\ref{Pconv}). Hence, conservation of total momentum in Eq.(\ref{Pconv}) allows us to write
\begin{equation}
\fl	\Delta\langle\mathbf{P}^{\textrm{sf}}_{\textrm{kin}}\rangle\simeq\langle n^{0}_{\downarrow}|\mathbf{P}^{Z}_{\parallel}+\mathbf{P}^{Z}_{\perp}+\mathbf{P}^{W}_{\parallel}+\mathbf{P}^{W}_{\perp}|n^{0}_{\downarrow}\rangle
	-\langle n^{0}_{\uparrow}|\mathbf{P}^{Z}_{\parallel}+\mathbf{P}^{Z}_{\perp}+\mathbf{P}^{W}_{\parallel}+\mathbf{P}^{W}_{\perp}|n^{0}_{\uparrow}\rangle.\label{Pkinfin}
\end{equation}
Finally, symmetry considerations imply that the two terms on the right-hand-side of Eq.(\ref{Pkinfin}) have equivalent strengths and opposite signs. Hence, it suffices to compute any of them. Correspondingly, there exists a net nuclear force of strength $\sim\omega_{1}\Delta\langle P^{\textrm{sf}}_{\textrm{kin}}\rangle$ all along the spin reversal process.
\begin{figure}[h]
\centering
	\includegraphics[height=7.2cm,width=8.5cm,clip]{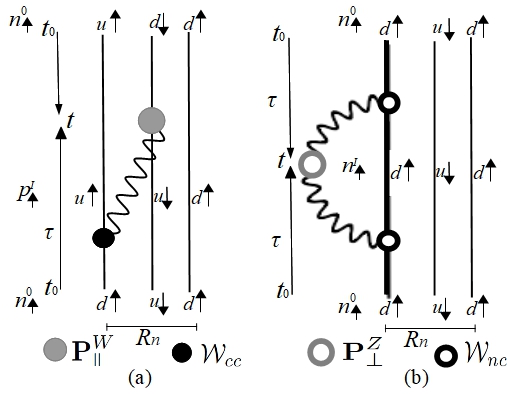}
	\caption{Diagrammatic representation of two of the processes contributing to Eq.(\ref{Ppor2}) for $\langle n^{0}_{\uparrow}|\mathbf{P}^{\textrm{W}}_{\parallel}|n^{0}_{\uparrow}\rangle$ [(a)] and to Eq.(\ref{Ppar2a}) for $\langle n^{0}_{\uparrow}|\mathbf{P}^{\textrm{Z}}_{\perp}|n^{0}_{\uparrow}\rangle$ [(b)]. Time runs along the vertical from $0$ up to the time of observation $t$ at which the momentum operators apply. Interaction vertices enter at the intermediate time $\tau$.}\label{fig1}
\end{figure}

In the calculation of Eq.(\ref{Pkinfin}) the spin dynamics can be considered adiabatic with respect to that of the intermediate states, since $\omega_{1}$ is much less than the transitions frequencies and decay rates of those states, i.e., the bosons' masses $M_{W,Z}$, and decay rates, $\Gamma_{Z,W}$, satisfying $M_{Z,W}\gg\Gamma_{Z,W}\gg\omega_{1}$.  Bearing this in mind and using time-dependent perturbation theory, we proceed to compute the longitudinal and transverse momenta of Z and W bosons.

Let us start by defining a non-perturbative Hamiltonian $H_{0}$ made of the neutron Hamiltonian, the Hamiltonians of the free boson fields and the effective Rabi Hamiltonian which causes the oscillations of the neutron spin, $H_{0}=H_{F}^{A}+H_{F}^{Z}+H_{F}^{W^{\pm}}+H_{n}+W^{R}_{EM}$. Correspondingly,  the nonperturbative time-evolution propagator is 
$\mathbb{U}_{0}(\tau)=\exp{(-i\tau\:H_{0})}$. At leading order in the perturbative potentials, $\mathcal{W}_{nc}$ and $\mathcal{W}_{cc}$, the expectation values of the longitudinal momenta 
of Z and W-bosons of Eq.(\ref{Pkinfin})  are computed at first order, while those of the transverse momenta are computed at second order. These are, respectively,
\begin{eqnarray}
\fl	\langle\mathbf{P}^{\textrm{Z,W}}_{\parallel}(t)\rangle&=
	2\textrm{Re}\langle n^{0}_{\downarrow}|\mathbb{U}_{0}^{\dagger}(t )\mathbf{P}^{\textrm{Z,W}}_{\parallel}
	\int _{0}^{t}-i\:d\tau e^{-\Gamma_{Z,W}(t-\tau)/2}\mathbb{U}_{0}(t-\tau)W_{nc,cc}\mathbb{U}_{0}(\tau )|n^{0}_{\downarrow}\rangle,\label{Ppar}\\
\fl	\langle\mathbf{P}^{\textrm{Z,W}}_{\perp}(t)\rangle&=\langle n^{0}_{\downarrow}|\int _{0}^{t}d\tau' 
	e^{-\Gamma_{Z,W}(t-\tau')/2}\mathbb{U}_{0}^{\dagger}(\tau' )W^{\dagger}_{nc,cc}
	\mathbb{U}_{0}^{\dagger}(t-\tau')\:\mathbf{P}^{\textrm{Z,W}}_{\perp}\nonumber\\
\fl	&\times\int _{0}^{t}d\tau e^{-\Gamma_{Z,W}(t-\tau)/2}\mathbb{U}_{0}(t-\tau)W_{nc,cc}\mathbb{U}_{0}(\tau )|n^{0}_{\downarrow}\rangle.\label{Pper}
\end{eqnarray} 
The variation of the kinetic momentum all along the spin-flip, $\Delta\langle\mathbf{P}_{\textrm{kin}}^{\textrm{sf}}\rangle$, comprises the difference  between the evaluation of the above 
equations for $t\rightarrow\pi/\omega_{1}$ (i.e., for $|n(\pi/\omega_{1})\rangle=|n^{0}_{\uparrow}\rangle$) and $t\rightarrow0^{+}$ with $t\Gamma_{Z,W}\gg1$ (i.e., for $|n(0)\rangle=|n^{0}_{\downarrow}\rangle$).  
The diagrams in Fig.\ref{fig1} depict two of the processes contributing to  $\langle n^{0}_{\uparrow}|\mathbf{P}^{\textrm{W}}_{\parallel}|n^{0}_{\uparrow}\rangle$ [(a)] 
and $\langle n^{0}_{\uparrow}|\mathbf{P}^{\textrm{Z}}_{\perp}|n^{0}_{\uparrow}\rangle$ [(b)]. In all the contributing intermediate states the spins of all the quarks are preserved by the interaction with Z and W-bosons. Hence, the intermediate states $\{|I\rangle\}$ in the diagrams are spin-up protons $|p^{I}_{\uparrow}\rangle$ for $\langle\mathbf{P}^{W}_{\parallel,\perp}\rangle$ and spin-up neutrons $|n^{I}_{\uparrow}\rangle$ for $\langle\mathbf{P}^{Z}_{\parallel,\perp}\rangle$.

As for the longitudinal momenta, the evaluation of Eq.(\ref{Ppar}) at $t=\pi/\omega_{1}$ yields, in each case,
\begin{eqnarray}
	\langle n^{0}_{\uparrow}|\mathbf{P}^{\textrm{Z}}_{\parallel}|n^{0}_{\uparrow}\rangle&=-2\textrm{Re}\sum_{I}\left\langle\frac{\langle n^{0}_{\uparrow}|\mathbf{P}^{\textrm{Z}}_{\parallel}|n^{I}_{\uparrow}\rangle\langle n^{I}_{\uparrow}|W_{nc}|n^{0}_{\uparrow}\rangle}{H^{Z}_{F}+\mathcal{E}_{I}-m_{n}-i\Gamma_{Z}/2}\right\rangle,\label{Ppor1}\\
	\langle n^{0}_{\uparrow}|\mathbf{P}^{\textrm{W}}_{\parallel}|n^{0}_{\uparrow}\rangle&=-2\textrm{Re}\sum_{I}\left\langle\frac{\langle n^{0}_{\uparrow}|\mathbf{P}^{\textrm{W}}_{\parallel}|p^{I}_{\uparrow}\rangle\langle p^{I}_{\uparrow}|W_{cc}|n^{0}_{\uparrow}\rangle}{H^{W}_{F}+\mathcal{E}_{I}-m_{n}-i\Gamma_{W}/2}\right\rangle,\label{Ppor2}
\end{eqnarray}
where the big brackets denote the expectation value of the boson field operators evaluated in the vacuum state, $m_{n}$ is the mass of the neutron at rest, and
 $\mathcal{E}_{I}$ are the energies of the intermediate states.  Next, it can be verified that, for the MIT-bag model, the overlap between the spatial wavefunction of the ground state and  those of excited states is  small.  On the other hand, in the HO model the excited sates exceed the nonrelativistic limit. Hence, it suffices hereafter to consider only the spin-up ground states of neutrons and protons in the sums over intermediate states, $|n^{0}_{\uparrow}\rangle$ and $|p^{0}_{\uparrow}\rangle$. Thus,  writing the transient ($t$) weak currents and chiral charge densities as 
\begin{eqnarray}
	\mathbf{j}_{nc}^{t}(\mathbf{y})&=\langle n^{0}_{\uparrow}|\sum_{i=1}^{3}\mathbf{j}^{i}_{nc}(\mathbf{y})|n^{0}_{\uparrow}\rangle,\quad	j_{nc}^{0,t}(\mathbf{x})=\langle n^{0}_{\uparrow}|\sum_{i=1}^{3}j_{nc}^{i,0}(\mathbf{x})|n^{0}_{\uparrow}\rangle\textrm{ for }\langle\mathbf{P}^{\textrm{Z}}_{\parallel}\rangle,\label{jnc}\\
	\mathbf{j}_{cc}^{t}(\mathbf{y})&=\langle p^{0}_{\uparrow}|\sum_{i=1}^{3}\mathbf{j}^{i}_{cc}(\mathbf{y})|n^{0}_{\uparrow}\rangle,\quad j_{cc}^{0,t}(\mathbf{x})=\langle p^{0}_{\uparrow}|\sum_{i=1}^{3}j_{cc}^{i,0}(\mathbf{x})|n^{0}_{\uparrow}\rangle\textrm{ for }\langle\mathbf{P}^{\textrm{W}}_{\parallel}\rangle,\label{jcc}
\end{eqnarray}
Eqs.(\ref{Ppor1}) and (\ref{Ppor2}) can be rewritten as
\begin{eqnarray}
\langle n^{0}_{\uparrow}|\mathbf{P}^{\textrm{Z}}_{\parallel}|n^{0}_{\uparrow}\rangle&=\frac{-g^{2}}{2\cos^{2}{\theta_{W}}}\textrm{Re}\int d^{3}x\int d^{3}y\left\langle j_{nc}^{0,t}(\mathbf{x})\mathbf{Z}(\mathbf{x})\frac{\mathbf{Z}^{\dagger}(\mathbf{y})\cdot\mathbf{j}_{nc}^{t\dagger}(\mathbf{y})}{H^{Z}_{F}}\right\rangle,\label{loparZ}\\
\langle n^{0}_{\uparrow}|\mathbf{P}^{\textrm{W}}_{\parallel}|n^{0}_{\uparrow}\rangle&=-g^{2}\textrm{Re}\int d^{3}x\int d^{3}y\left\langle j_{cc}^{0,t}(\mathbf{x})W^{+}(\mathbf{x})\frac{\mathbf{W}^{-}(\mathbf{y})\cdot\mathbf{j}_{cc}^{t\dagger}(\mathbf{y})}{H^{W}_{F}}\right\rangle.\label{loparW}
\end{eqnarray}

As for the transverse momenta, the evaluation of Eq.(\ref{Pper}) at $t=\pi/\omega_{1}$ yields
\begin{eqnarray}
\langle n^{0}_{\uparrow}|\mathbf{P}^{\textrm{Z}}_{\perp}|n^{0}_{\uparrow}\rangle&=\sum_{I}\left\langle\frac{\langle n^{0}_{\uparrow}|W^{\dagger}_{nc}|n^{I}_{\uparrow}\rangle\mathbf{P}^{\textrm{Z}}_{\perp}\langle n^{I}_{\uparrow}|W_{nc}|n^{0}_{\uparrow}\rangle}{(H^{Z}_{F}+\mathcal{E}_{I}-m_{n})^{2}+\Gamma_{Z}^{2}}\right\rangle,\label{Ppar2a}\\
\langle n^{0}_{\uparrow}|\mathbf{P}^{\textrm{W}}_{\perp}|n^{0}_{\uparrow}\rangle&=\sum_{I}\left\langle\frac{\langle n^{0}_{\uparrow}|W_{cc}^{\dagger}|p^{I}_{\uparrow}\rangle\mathbf{P}^{\textrm{W}}_{\perp}\langle p^{I}_{\uparrow}|W_{cc}|n^{0}_{\uparrow}\rangle}{(H^{W}_{F}+\mathcal{E}_{I}-m_{n})^{2}+\Gamma_{W}^{2}}\right\rangle,\label{Ppar2b}
\end{eqnarray}
which, restricting the sums over intermediate nucleon states to  $|n^{I}_{\uparrow}\rangle=|n^{0}_{\uparrow}\rangle$, $|p^{I}_{\uparrow}\rangle=|p^{0}_{\uparrow}\rangle$ in either case, give
\begin{eqnarray}
\fl\langle n^{0}_{\uparrow}|\mathbf{P}^{\textrm{Z}}_{\perp}|n^{0}_{\uparrow}\rangle&=\frac{g^{2}}{2\cos^{2}{\theta_{W}}}\int d^{3}x\int d^{3}y\left\langle\frac{j_{nc}^{0,t}(\mathbf{x})Z_{0}(\mathbf{x})}{H^{Z}_{F}}\mathbf{P}_{\perp}^{Z}\frac{\mathbf{Z}(\mathbf{y})\cdot\mathbf{j}_{nc}^{t\dagger}(\mathbf{y})}{H^{Z}_{F}}\right\rangle,\label{loperZ}\\
\fl\langle n^{0}_{\uparrow}|\mathbf{P}^{\textrm{W}}_{\perp}|n^{0}_{\uparrow}\rangle&=g^{2}\int d^{3}x\int d^{3}y
\left\langle\frac{j_{cc}^{0,t}(\mathbf{x})W^{+}_{0}(\mathbf{x})}{H^{W}_{F}}\mathbf{P}_{\perp}^{W}\frac{\mathbf{W}^{-}(\mathbf{y})\cdot\mathbf{j}_{cc}^{t\dagger}(\mathbf{y})}{H^{W}_{F}}\right\rangle.\label{loperW}
\end{eqnarray}
In Eqs.(\ref{loparZ}), (\ref{loparW}), (\ref{loperZ}), (\ref{loperW}) the transient weak currents are directed along the neutron spin.  Finally, expanding the quadratic vacuum fluctuations of the boson fields as sums over normal modes, and adding up longitudinal and transverse contributions we arrive at
\begin{eqnarray}
\fl\langle n^{0}_{\uparrow}|\mathbf{P}^{\textrm{Z}}_{\parallel}+\mathbf{P}^{\textrm{Z}}_{\perp}|n^{0}_{\uparrow}\rangle&=\frac{-g^{2}/4}{\cos^{2}{\theta_{W}}}\textrm{Re}\int d^{3}x\int d^{3}y\int\frac{d^{3}q}{(2\pi)^{3}}j_{nc}^{0,t}(\mathbf{x})\frac{e^{-i\mathbf{q}\cdot(\mathbf{x}-\mathbf{y})}}{q^{2}+M^{2}_{Z}}\:\mathbf{j}^{t\dagger}_{nc}(\mathbf{y}),\label{final_formZ}\\
\fl\langle n^{0}_{\uparrow}|\mathbf{P}^{\textrm{W}}_{\parallel}+\mathbf{P}^{\textrm{W}}_{\perp}|n^{0}_{\uparrow}\rangle&=\frac{-g^{2}}{2}\textrm{Re}\int d^{3}x\int d^{3}y\int\frac{d^{3}q}{(2\pi)^{3}}j_{cc}^{0,t}(\mathbf{x})\frac{e^{-i\mathbf{q}\cdot(\mathbf{x}-\mathbf{y})}}{q^{2}+M^{2}_{W}}\:\mathbf{j}^{t\dagger}_{cc}(\mathbf{y}).\label{final_formW}
\end{eqnarray}
From Eqs.(\ref{final_formZ}) and (\ref{final_formW})  we interpret that the variation of the momenta of Z and W bosons results from the interaction between the transient chiral charge densities $j_{nc,cc}^{0,t}$ and weak currents  $\mathbf{j}_{nc,cc}^{t}$ of the neutron's quarks. In virtue of Eq.(\ref{Pkinfin}) an equivalent variation of the kinetic momentum is carried off by the neutron in the opposite direction. Since the weak currents reverse direction when the quarks' spins are inverted, the spin-flip  of the neutron results in a net variation of its total kinetic momentum, and explains the equivalent contribution of the two terms in Eq.(\ref{Pkinfin}).  The 
finiteness of the calculation is guaranteed by the short-range of the weak nuclear forces and the confinement of quarks. Hence, the momentum integrals in the above equations scales as $\sim e^{-|x-y|M_{V}}/(xyM_{V})$. Also worth noting is that Eqs.(\ref{final_formZ}) and (\ref{final_formW})  are the EW analogue of the longitudinal momentum of the EM field sourced by a chiral molecule, $\mathbf{E}\times\mathbf{B}$, where the electrostatic field is sourced by the electric charges and the magnetic field is created by its helical currents  \cite{JPCMDonaire}.
Finally, including in those equations the complete expressions of the transient currents and charges, integrating in $q$ and in one of the spatial coordinates, and   performing the sum of the momenta in Eq.(\ref{Pkinfin}), we obtain
\begin{equation}
\fl\Delta\langle\mathbf{P}^{\textrm{sf}}_{\textrm{kin}}\rangle\simeq\hat{\mathbf{z}}\frac{-20\pi G_{F}}{\sqrt{2}R_{n}^{3}}\mathcal{F}_{n},\textrm{ with }
\mathcal{F}_{n}=\int d\tilde{r}\:\tilde{r}^{2}\left[F^{2}(\tilde{r})+G^{2}(\tilde{r})\right]\left[F^{2}(\tilde{r})-\frac{G^{2}(\tilde{r})}{3}\right].\label{result1}
\end{equation}
An outline of the technicalities in the calculation of the
transient weak currents and chiral charge densities, together with the expansions of the quadratic vacuum field fluctuations in normal modes and with the calculation of the integrals, can be found in \ref{apptech}. In Eq.(\ref{result1}) the factor $\mathcal{F}_{n}$ is a dimensionless integral which depends on the upper and lower components of the quarks' ground state wavefunction, $\psi^{0,\uparrow}_{u,d}\approx\bigl(F(\tilde{r})\uparrow\:\:\:iG(\tilde{r})\hat{\mathbf{r}}\cdot\boldsymbol{\sigma}\uparrow\bigr)$, with $\tilde{r}=r/R_{n}$.  In particular, for the MIT-bag relativistic model, the total variation of the neutron velocity along the spin-flip is $13\:$m/s,  while for the nonrelativistic harmonic-oscillator model it is  $7\:$m/s approximately. 

Lastly, the variation of the kinetic momentum during the spin reversal process led by the RF field of strength $B_{1}$ is driven by a net 
 EW force along the spin direction, $\hat{\bf{z}}$. From the expression in Eq.(\ref{primera}) it can be shown that this force is $\omega_{1}\sin{(\omega_{1}t)}\Delta\langle P^{\textrm{sf}}_{\textrm{kin}}\rangle/2$. 
An outline of its computation  is given in \ref{appforce}.

\section{Conclusions and discussion}\label{sec:conclus}	

In this final Section we summarize the results and offer additional interpretations. We outline an experimental proposal for the verification of the effect and explain why it has not yet been observed in previous experimental setups. Lastly, we make a critical analysis on the accuracy of our quantitative result and on the theoretical assumptions considered in its estimate.

\subsection{Results interpretation}
	
Net internal weak nuclear forces have been found to operate along the spin-reversal of a
polarized neutron. On physical grounds, they result from the chiral nature of the EW
interactions between quarks, together with the time-reversal symmetry breaking provided by the
spin polarization. The apparent contradiction with the action-reaction principle is resolved by
considering the momentum transferred to the fields of the bosons which mediate the interaction.
In this respect, the effect is analogous to that which causes the directionality of the beta decay
along the nuclear spin. The difference being that here the recoil of the neutron is against the
fields of the EW bosons, whereas in the beta decay it is against actual neutrinos and electrons. Using the simplest hadron models, namely the MIT-bag  and the harmonic-oscillator models, it is found a velocity variation of the order of $\sim$~m/s. 
In regards to energetics, the acceleration of the neutron by no means implies the extraction of
energy from the quantum vacuum. On the contrary, a calculation analogous to that of
Ref.\cite{JPCMDonaire} for the charges bound in a magnetochiral molecule but for the quarks
bound in a neutron in this case, reveals that the kinetic energy gained by the neutron
corresponds to corrections to its EW self-energy. Using the nonrelativistic harmonic-oscillator
model it is shown in \ref{energetics_source} that that kinetic energy results from the adiabatic
variation of the magnetization of the neutron. Hence, for a polarized neutron starting at rest, its
kinetic energy after the spin-flip goes like
$\sim\langle\mathbf{S}_{n}\rangle\cdot\langle\mathbf{S}_{n}\rangle$, with $\mathbf{S}_{n}$
being the neutron spin operator. Thus, as for the case of a magnetochiral molecule
\cite{JPCMDonaire}, the kinetic energy is provided by the sources of the external magnetic fields
that cause the spin reversal.

\subsection{Experimental verification}

 Given a theoretical prediction for the velocity variation as large as $\sim$m/s, it is reasonable to wonder how such a strong effect has not yet been observed in current ultracold neutron (UCN) experiments. Hence, neutron polarization, spin-flips and storage of slow neutrons are usual techniques in UCN experiments. However, there exist several issues in the typical experimental setups which prevent the observation of the effect. More specifically, the following features are to be held in any experimental setup in order to detect the EW-induced acceleration of the neutrons. In the first place, the spin-reversal process must be performed upon a polarized ensemble of neutrons, and not upon an incoherent mixture of spin up and spin down species. Hence, our calculations predict null momentum transfer for the case of an incoherent mixture of spin up and spin down neutron. Second, the spin reversal process must be coherent. This is the reason for using a combination of an RF field together with a static and uniform magnetic field. Note also that the adiabaticity of the spin flip process with respect to the internal dynamics of the neutron is common to any setup where the polarization is driven by magnetic fields. Third, the measurement of the initial velocity of the neutrons, the spin-reversal process, and the measurement of the final velocity must be performed within the area where the static magnetic field is approximately uniform. This must be so in order to avoid any velocity variation  due to gradients of the static field. Finally, the flight of the neutrons along the whole process must be free of collisions which affect anyhow the helicity of the neutrons. 
 
 All the aforementioned features are not generally met in UCN experiments. For instance, in UCN experiments for the detection of a neutron electric dipole moment, cosmic fields, etc., the storage of slow neutrons is induced by spin-flip in a macroscopic chamber --see,  eg., Ref.\cite{PhysicaB}. Since the maximal velocity for the stored neutrons is typically a few m/s, the velocity variation estimated in Sec.\ref{EWmomentum} would cause a complete leak of neutrons from the chamber, which is not the case. Typically, the volume of the chambers is about $\sim 10^3$cm$^{3}$, and the spin-flip is driven by the Ramsey method in time intervals of the order of tens of  seconds and along a direction transverse to that of the net propagation of the neutron beam. Thus, during that time the neutrons experience a number of reflections off the chamber walls which cause the corresponding variations in their helicities and momenta. Since the effect sought in this article relies on the conservation of total linear momentum and the coherent variation of helicity, the randomness of the variations caused by multiple reflections results in zero velocity variation in average, and hence no leak of neutrons from the chamber is to be expected. Also, in other experiments which use longitudinal spin polarization in combination with fast adiabatic spin flip --see, eg., Ref.\cite{PRL_Induzierte}, the UCN beam consists in an incoherent admixture of $50\%$  spin up and $50\%$ spin down neutrons. The EW-induced velocity variation corresponding to such a statical ensemble would be identically zero by symmetry considerations, since the contribution of each polarization species have opposite signs. 
 
 Instead, our experimental proposal is based on current neutron spectrometers' configurations  \cite{Rafik}. Specifically, we aim at measuring the difference in the time of flight between two neutron beams which are polarized along their propagation trajectory with opposite helicities. Each neutron beam would fly freely while being spin-flipped though a dedicated RF $\pi$-flipper. The helicity of the neutrons would change coherently by effect of spin rotation and the corresponding variations of the kinetic momentum would add up adiabatically along the way. Since the latter would be of equal strength and opposite sign for each helicity species, that would cause a difference between the times of flight of each beam in their way to the detector.

\subsection{Uncertainty sources in the theoretical modelling}

In principle, the scale dependence of the variation of the kinetic momentum in Eq.(\ref{result1}), $G_{F}/R_{n}^{3}$, could have been worked out by dimensional analysis. The EW interaction enters quadratically in the calculation with coupling constant $g$ and length-scale $1/M_{W,Z}$. Hence, the linear dependence in $G_{F}=\sqrt{2}g^{2}/8M^{2}_{W}$. On the other hand, the resultant kinetic energy has been proved to be part of the EW self-energy of the bound quarks. Hence, the inverse proportion with the confinement volume, $1/R_{n}^{3}$.

However, the magnitude of the estimated velocity depends strongly on the dimensionless factors of  Eq.(\ref{result1}). Thus,  some comments are in order in regards to the  accuracy of our estimate and its dependence on the hadron models. In particular, two are the main reasons to believe that the magnitude of the velocity so computed is overestimated. Namely, the inaccuracy in the quark wavefunction of excited states, and the uncertainty in the distribution of the internal angular momentum of the nucleon.  

In the first place, from  Eq.(\ref{result1}) we see that the estimate is highly sensitive to the spatial quark distribution through the wavefunctions in $\mathcal{F}_{n}$ and the effective nucleon radius $R_{n}$. More importantly, it has been argued that the major contribution to the sum over intermediate states in Eqs.(\ref{Ppor1}), (\ref{Ppor2}), (\ref{Ppar2a}), (\ref{Ppar2b}) comes from intermediate nucleon states in which all the quarks remain in their ground state. This is so because the overlap between the wavefunctions of the excited and the ground states is mall in the spatial integrals. The same was found in the computation of the QCD self-energy of the nucleons in the seminal papers of the MIT-bag model \cite{MIT_bag1}. However, this seems to be specific to the simple MIT-bag model that we use, and there is no reason to believe that the contribution of excited states must be negligible. Hence, 
any  refinement upon our simplified models of the sort of chiral bag models, Goldstone-bosons potentials, one-gluon-exchange potentials, etc. \cite{Rujula_Cornell} must be considered.  

Second, the sought effect relies strongly on the variation of the spin of the  valence quarks. Thus, the optimization of the effect goes through the maximization of their spin variation.  
In this article we have considered the simplest assumption, namely, that the nucleon spin results entirely from the addition of the spins of the three valence quarks which, in the ground state, present zero orbital angular momentum. However, from the advent of the \emph{spin crisis} \cite{spincrisis} it is recognized that the total spin of the nucleons may have different contributions from gluons, pion clouds, sea quarks, etc., and even the valence quarks themselves may present a net orbital angular momentum. All this could modify the wavefunctions of the quarks which enter $\mathcal{F}_{n}$ in Eq.(\ref{result1}) and, more importantly, it would diminish the net variation of the spin of the quarks along the spin-flip of the neutron, thus reducing the variation of its kinetic momentum. 

All in all, important deviation from our simplified model are to be expected, and all of them leading to underestimate the sought effect. In this respect, a velocity variation of the order of m/s can be considered as an upper bound estimate. More realistic models for nucleons are needed to improve the accuracy of the estimate \cite{further}. Nonetheless, even if the effect turns out to be  some orders of magnitude weaker in reality, symmetry considerations prevents it from being null. Hence, our finding may be used to probe the structure and the angular momentum content of the nucleons, and as a test for hadron models, employing low energy experimental methods.

\ack
We thank  B. van Tiggelen, G. Rikken and D.R. Entem for fruitful discussions on the fundamentals of the phenomenon, R. Ballou and G. Pignol for enlightening discussions on the experimental techniques for its observation, and A. Cano for his hospitality at Institut N\'eel. Financial support from the NextGenerationEU funds, through the Spanish Ministerio de Universidades, is acknowledged.

\appendix

\section{Hadron confinement models}\label{app1}

For simplicity, two effective models of hadron confinement have been considered in the article for the numerical estimations, namely, that of the MIT-bag  for valence and relativistic quarks 
\cite{MIT_bag1}, and the harmonic-oscillator model for constituent and nonrelativistic quarks \cite{Isgur}. Formally, their most remarkable difference 
is that only the latter allows for an explicit separation between internal and external degrees of freedom. In fact, as for the original MIT-bag model, it is well known that if fails to preserve the chiral symmetry and, more importantly, it hinders a consistent treatment of the nucleon center of mass. Indeed, its Hamiltonian does not  fit into that of Eq.(\ref{Ho}). As for the harmonic-oscillator  model, it is known that if fails to reproduce the spectrum of excited states. Nonetheless, for our purposes here, they both are sufficient to describe the spatial wavefunctions of the valence and the constituent quarks, respectively.   

\subsection{Relativistic MIT-bag model}\label{MIT-bag}

In the MIT-bag model (B),  the Lagrangian density of Ref.\cite{MIT_bag1}  for the three valence quarks bound in the neutron is, in the massless approximation,
\begin{eqnarray}
	\mathcal{L}_{0}^{\textrm{B}}=\sum_{j=1}^{3}\Bigl\{\Bigl[\frac{i}{2}\left(\bar{\psi}_{j}\gamma^{\mu}\partial_{\mu}\psi_{j}-\left(\partial_{\mu}\bar{\psi}_{j}\right)\gamma^{\mu}\partial_{\mu}\psi_{j}\right)-B\Bigr]\Theta(|\mathbf{r}_{j}-\mathbf{R}|-R_{n})\nonumber\\
	-\frac{1}{2}\bar{\psi}_{j}\psi_{j}\delta(|\mathbf{r}_{j}-\mathbf{R}|-R_{n})\Bigr\}.\label{LagrangeB}
\end{eqnarray} 
From this Lagrangian, one derives the Hamiltonian, 
\begin{equation}
	H_{n}^{\textrm{B}}=-i\sum_{j=1}^{3}\int d^{3}r_{j}\bar{\psi}_{j}\gamma^{k}\nabla_{k}\psi_{j}\Theta(|\mathbf{r}_{j}-\mathbf{R}|-R_{n}),\label{HB}
\end{equation} 
together with the boundary conditions, 
\begin{equation}
	-i\hat{\mathbf{R}}_{n}\cdot\boldsymbol{\gamma}\psi_{j}|_{\partial\Omega_{n}}=\psi_{j}|_{\partial\Omega_{n}},
	\quad \hat{\mathbf{R}}_{n}\cdot\boldsymbol{\nabla}(\bar{\psi}_{j}\psi_{j})|_{\partial\Omega_{n}}=-2B.\label{B.C.}
\end{equation}
In the above equations the subscript $j$ runs over the three valence quarks, with position vectors $\mathbf{r}_{j}$, $\mathbf{R}$ being the center of the 
bag;  $\{\psi_{j}=\psi(\mathbf{r}_{j})\}$ are Dirac spinors;  $R_{n}$ is the static radius of the spherical 
cavity with outwards unitary vector $\hat{\mathbf{R}}_{n}$ at the surface $\partial\Omega_{n}$. The linear boundary condition in Eq.(\ref{B.C.}) implies no current flow 
through the surface, whereas the nonlinear one implies the  balance between the outwards pressure of the quarks and the inward QCD vacuum pressure $B$ at the surface.  

The components of the spinor eigenfunctions $\{\psi(\mathbf{r}_{j})\}$ can be split in upper and lower components, each one labeled with 
four common numbers, $n$, $\kappa$,  $J$ and $J_{3}$; and different orbital angular momentum numbers, $l$ and $l'$. 

\[
\psi(\mathbf{r})=\left( \begin{array}{c}F_{n\kappa}(r)\mathcal{Y}_{l,1/2}^{J,J_{3}}(\hat{\mathbf{r}})\\\\i\:G_{n\kappa}(r)\mathcal{Y}_{l',1/2}^{J,J_{3}}(\hat{\mathbf{r}})\end{array} \right)=
\mathcal{N}_{n,\kappa}\left( \begin{array}{c}j_{l}(\mathcal{E}_{n,\kappa}r)\mathcal{Y}_{l,1/2}^{J,J_{3}}(\hat{\mathbf{r}})\\\\(-1)^{\eta}i\:j_{l'}(\mathcal{E}_{n,\kappa}r)\mathcal{Y}_{l',1/2}^{J,J_{3}}(\hat{\mathbf{r}})\end{array} 
\right).
\]
In these equations $\mathcal{N}_{n,\kappa}$ is a normalization constant, $j_{l^{(')}}$ are spherical Bessel functions, $\mathcal{Y}_{l^{(')},1/2}^{J,J_{3}}$ are spinor spherical harmonics, and we use a representation in which Dirac's matrices are given by 
\[
\gamma_{0}=
\left( {\begin{array}{cc}
		\mathbb{I}_{2} & 0\\       0&-\mathbb{I}_{2}\      \end{array} } \right),\quad \gamma_{5}=
\left( {\begin{array}{cc}
		0 & \mathbb{I}_{2}\\   \mathbb{I}_{2} & 0\      \end{array} } \right),\quad \gamma^{i}=
\left( {\begin{array}{cc}
		0 & \sigma_{i}\\       -\sigma_{i} &0\      \end{array} } \right).
\]
$\mathcal{E}_{n,\kappa}$ is the $n$-eigenenergy value, i.e., the $n^{th}$ solution of a transcendental 
equation for certain eigenvalue $\kappa$ of the operator $\mathcal{K}=\gamma_{0}(\boldsymbol{\Sigma}\cdot\mathbf{L}+\mathbb{I})=
\gamma_{0}(\boldsymbol{\Sigma}\cdot\mathbf{J}-\mathbb{I}/2)$, which derives from 
the equations of the boundary coniditions. For masseless quarks, 
\begin{equation}
\fl	j_{l}(\mathcal{E}_{n,\kappa}R_{n})=(-1)^{\eta}j_{l+1}(\mathcal{E}_{n,\kappa}R_{n}),\textrm{ with }\eta=0\textrm{ for }\kappa<0, \eta=1\textrm{ for }\kappa>0.
\end{equation}
The number $\kappa$ may take positive and negative values which depend on the total angular momentum $J$, $\kappa=\pm(J+1/2)$. The orbital angular momenta are $l=-(\kappa+1)$ and $l'=-\kappa$ for $\kappa<0$; and 
$l=\kappa$, $l'=\kappa-1$ for $\kappa>0$. In particular, the wavefunctions of the first two levels with $J=1/2$ and opposite parity (i.e., $1S_{1/2}$ and   $1P_{1/2}$ states) are  
\begin{equation}
\fl\psi_{1,-1}^{s}(\mathbf{r})=\mathcal{N}_{1,-1}\left( \begin{array}{c}j_{0}(\mathcal{E}_{1,-1}r)\:\chi_{s}\nonumber\\\\
	i\:j_{1}(\mathcal{E}_{1,-1}r)\hat{\mathbf{r}}\cdot\boldsymbol{\sigma}\:\chi_{s}\end{array} \right),\nonumber
\psi_{1,+1}^{s}(\mathbf{r})=\mathcal{N}_{1,+1}\left( \begin{array}{c}j_{1}(\mathcal{E}_{1,+1}r)\hat{\mathbf{r}}\cdot\boldsymbol{\sigma}\:\chi_{s}\nonumber\\\\
	-ij_{0}(\mathcal{E}_{1,+1}r)\:\chi_{s}\end{array} \right),\nonumber
\end{equation}
where $\chi_{s}$ are unit two-component spinors, $s=+$ for spin-up $\uparrow$, and $s=-$ for-spin down $\downarrow$; and the eigenenergies are $\mathcal{E}_{1,-1}=2.04/R_{n}$,  $\mathcal{E}_{1,+1}=3.81/R_{n}$. It is $\psi^{s}_{1,-1}(\mathbf{r}_{j})$ the one to be identified with $\psi_{f_{j}}^{0,s_{j}}$ for any valence quark $j$ of flavour $f_{j}$ and spin $s_{j}$, in the wavefunction of the neutron ground state --see below.

Finally, in order to fit the nucleon radius for any combination of wavefunctions of the three valence quarks, one requires the balance between the internal  pressure exerted  by the quarks and the external vacuum pressure, $-B$, at the surface of the bag. This implies the boundary condition
\begin{equation}
	\frac{d\mathcal{E}_{tot}}{dr}|_{\partial\Omega_{n}}=\frac{d}{dr}\left[\frac{4\pi}{3}Br^{3}+\sum_{j=1}^{3}\mathcal{E}^{j}R_{n}/r-z_{0}/r\right]\Bigr|_{r=R_{n}}=0,
\end{equation} 
where $\mathcal{E}^{j}$ is the energy of the wavefunction of quark $j$ and $z_{0}$ is  a numerical parameter of value 1.5 approximately which weights the contribution of an additional QCD vacuum energy needed to fit the mass spectrum together with the value of $B^{1/4}\approx 145\:$MeV \cite{MIT_bag2_Greiner,MIT_bag1}.  Note that the normalization constants  $\mathcal{N}_{n,\kappa}$ of each quark's wavefunction depend on the radius of the whole nucleus computed this way.

\subsection{Nonrelativistic harmonic oscillator model}\label{gauss}

As for the nonrelativistic harmonic oscillator (HO) model, its Hamiltonian is
\begin{equation}
	H_{n}^{\textrm{HO}}=\sum_{j=1}^{3}\int d^{3}r_{j}\psi^{\dagger}_{j}\Bigl[-\frac{\nabla^{2}}{2m}+\frac{m\omega^{2}}{4}\sum_{k<j}^{3}\int d^{3}r_{k}
	\:\psi^{\dagger}_{k}|\mathbf{r}_{j}-\mathbf{r}_{k}|^{2}\psi_{k}\Bigr]\psi_{j},\label{HHO}
\end{equation}
where the masses $m$ and the oscillator frequencies $\omega$ are equivalent for the three constituent quarks \footnote{For consistency, the interaction Hamiltonians with the gauge fields are also to be modified in the nonrelativistic limit. However, since we are only interested in the spatial wavefunction of the nucleon, it suffices to consider this limit upon $H_{n}^{\textrm{HO}}$ only.}. In the nonrelativistic limit, the lower component of the Dirac spinors is negligible, and the spin-space eigenfunctions of the quarks can be approximated by tensor products of complex scalar functions and two-component spinors. Further, the above Hamiltonian is separable in  terms of Jacobi's coordinates. 

In any quantum perturbative process the  EW interactions enter at the location of one of the constituent quarks at a time. We refer to it as \emph{active quark} and denote its position vector by $\mathbf{r}_{A}$. The  positions of the remaining quarks, i.e., the \emph{spectators}, will be denoted by $\mathbf{r}_{1}$,  $\mathbf{r}_{2}$ in the laboratory frame. Their corresponding conjugate momenta are $\mathbf{p}_{A}$, $\mathbf{p}_{1}$ and $\mathbf{p}_{2}$, respectively. Note that the physical meaning of the conjugate momenta of this Hamiltonian is that of kinetic momenta. The Jacobi position vectors and corresponding conjugate momenta read \cite{JHEP,Coince}, 
\begin{eqnarray}
	\mathbf{R}&=\frac{\mathbf{r}_{1}+\mathbf{r}_{2}+\mathbf{r}_{A}}{3},\quad\mathbf{r}_{\rho}=\mathbf{r}_{2}-\mathbf{r}_{1},\quad\mathbf{r}_{\lambda}=\mathbf{r}_{A}-\frac{\mathbf{r}_{1}+\mathbf{r}_{2}}{2},\nonumber\\
	\mathbf{p}&=\mathbf{p}_{1}+\mathbf{p}_{2}+\mathbf{p}_{A},\:\:\mathbf{p}_{\rho}=\frac{\mathbf{p}_{2}-\mathbf{p}_{1}}{2},\:\:\mathbf{p}_{\lambda}=\frac{2\mathbf{p}_{A}-\mathbf{p}_{1}-\mathbf{p}_{2}}{3},\nonumber
\end{eqnarray}
where $\mathbf{R}$ is the centre of mass of the nucleon and $\mathbf{p}$ its conjugate momentum.  As for the active quark, its position vector  $\mathbf{r}_{A}$ reads, in terms of Jacobi's coordinates, $\mathbf{r}_{A}=\mathbf{R}+2\mathbf{r}_{\lambda}/3$. Likewise, in terms of Jacobi's coordinates and momenta, the Hamiltonian is
\begin{equation}
	H_{n}^{\textrm{HO}}=p^{2}/6m+p_{\rho}^{2}/m + 3p_{\lambda}^{2}/4m + m\omega^{2}(r_{\rho}^{2}/4+r_{\lambda}^{2}/3).\label{Hint}
\end{equation}
The eigenfunctions in terms of Jacobi's coordinates reduce to the product of the wavefunctions of two independent and isotropic harmonic oscillators times plane wavefunctions for the free motion of the center of mass. In particular, the wavefunction for the ground state of the neutron with total conjugate momentum $\mathbf{p}$ is 
\begin{equation}
	\Psi^{0,s}_{n}(\mathbf{R},\mathbf{r}_{\lambda},\mathbf{r}_{\rho})=\frac{e^{-i\mathbf{p}\cdot\mathbf{R}}}{\sqrt{\mathcal{V}}}\frac{\beta^{3}}{3^{3/4}\pi^{3/2}}e^{-\beta^{2}r_{\lambda}^{2}/3}e^{-\beta^{2}r_{\rho}^{2}/4}\chi_{s},\label{wavef}
\end{equation}
with $\beta=\sqrt{m\omega}$, $\mathcal{V}$ being an infinite volume, and  $\chi_{s}$ a unit two-component spinor. Using the above formula for the calculation of $\Delta\langle\mathbf{P}_{\textrm{kin}}^{\textrm{sf}}\rangle$, integrating in $\mathbf{R}$ and $\mathbf{r}_{\rho}$, getting back to the  coordinates of the laboratory frame with $\tilde{x}\equiv 2r_{\lambda}/3R_{n}$, and taking $R_{n}=2\beta^{-1}\sqrt{2/3\pi}$ as the mean distance of the active quark to the center of the neutron, we find out $F(\tilde{x})=(2/\pi)^{3/2}e^{-2\tilde{x}^{2}/\pi}$ in the expression for $\mathcal{F}_{n}$ in Eq.(\ref{result1}). Note also that, for consistency with the actual values  of the mass and the radius of the neutron, $\omega\approx0.4\:m$, which implies that excited states would be inconsistent with the nonrelativistic assumption and thus are to be discarded.

\section{Technical details in the calculation of $\Delta\langle\mathbf{P}_{\textrm{kin}}^{\textrm{sf}}\rangle$}\label{apptech}

In this Appendix we outline the main details on the evaluation of the equations for $\langle n^{0}_{\uparrow}|\mathbf{P}^{\textrm{Z,W}}_{\parallel,\perp}|n^{0}_{\uparrow}\rangle$ in Section \ref{EWmomentum}.

In the first place, for the computation of the transient weak current and chiral charge densities we use the neutron wavefunction of $\Psi^{0}_{n\uparrow}$ in equation (\ref{Psinup}) with all the spins inverted (and likewise for the proton with the replacement $u\leftrightarrow d$) and the ground state wavefunctions for quarks given in \ref{MIT-bag} for the MIT-bag model. As for the densities of chiral charges that appear in Eqs.(\ref{loparZ}),(\ref{loparW}),(\ref{loperZ}),(\ref{loperW}), using the definitions of equations (\ref{jnc}) and (\ref{jcc}) we may write 
\begin{eqnarray}
\fl\frac{g}{2\cos{\theta_{W}}}j_{nc,v}^{0,t}(\mathbf{x})\:\mathbf{Z}(\mathbf{x})&=\frac{g}{2\cos{\theta_{W}}}(g_{v}^{u}+2g_{v}^{d})\psi^{s\dagger}_{1,-1}(\mathbf{x})\psi^{s'}_{1,-1}(\mathbf{x})\mathbf{Z}(\mathbf{x})\delta_{ss'},
	\label{momdenZ}\\
\fl	\frac{g}{\sqrt{2}}j_{cc,v}^{0,t}(\mathbf{x})\mathbf{W}^{+}(\mathbf{x})&=\frac{g}{2\sqrt{2}}\psi^{s\dagger}_{1,-1}(\mathbf{x})\psi^{s'}_{1,-1}(\mathbf{x})\mathbf{W}^{+}(\mathbf{x})\delta_{ss'}.\label{momdenW}
\end{eqnarray} 
Note that the chiral charge densities are only vector-like ($v$), $j_{(nc,cc),v}^{0,t}$, they are accompanied by delta functions of spin conservation, $\delta_{ss'}$, and their signs do not depend on the spin orientation. Using the nomenclature of the MIT-bag wavefunctions, $|\psi^{s}_{1,-1}(\mathbf{x})|^{2}=F_{1-1}^{2}(x)+G_{1-1}^{2}(x)$.  As for the transient weak currents, they are axial currents which also preserve the spin and go along the neutron spin, i.e., along $\hat{\mathbf{z}}$, 
\begin{eqnarray}
	\hat{\mathbf{z}}\cdot\mathbf{j}_{nc}^{t\dagger}(\mathbf{y})&=-\frac{4g_{a}^{d}-g_{a}^{u}}{3}\bar{\psi}^{s'}_{1,-1}(\mathbf{y})\gamma^{3}\gamma_{5}\psi^{s}_{1,-1}(\mathbf{y})\delta_{ss'},\nonumber\\
	\hat{\mathbf{z}}\cdot\mathbf{j}_{cc}^{t\dagger}(\mathbf{y})&=-\frac{5}{6}\bar{\psi}^{s'}_{1,-1}(\mathbf{y})\gamma^{3}\gamma_{5}\psi^{s}_{1,-1}(\mathbf{y})\delta_{ss'}.\label{corrientes}
\end{eqnarray}
These currents are the result of the addition of the contributions of each active quark with position vector $\mathbf{y}$, whose signs do depend on the spin orientation, $+$ for $\uparrow$ and $-$ for $\downarrow$, $\bar{\psi}^{s}_{1,-1}(\mathbf{y})\gamma^{3}\gamma_{5}\psi^{s}_{1,-1}(\mathbf{y})=[F_{1-1}^{2}(y)+\cos{2\theta_{y}}\:G_{1-1}^{2}(y)](\delta_{s+}-\delta_{s-})$ \footnote{In Eqs. (\ref{momdenZ})-(\ref{corrientes}), the spatial wavefunctions $\psi^{s(')}_{1,-1}(\mathbf{x,y})$ refer to those of the individual valence quarks.  They are common to all the quarks in the states $|n^{0}_{\uparrow\downarrow}\rangle$, $|p^{0}_{\uparrow\downarrow}\rangle$, but for their spin. The net result of the spin difference on the evaluation of the axial currents is   the multiplicative prefactors which account for the combination of up and down quarks in each of the three-quark components of those states. Thus, the remaining factor depends only on the net spin of the initial state,  i.e., $|n^{0}_{\uparrow}\rangle$ in our case. Hence, it must be understood that $s=+$ in all those equations.}. The prefactors of the above equations account for the combination of the different quark spins and weights of each of the components of $|n^{0}_{\uparrow}\rangle$ and $|p^{0}_{\uparrow}\rangle$.  It is worth noting that the combination of transient vector-like charges and transient axial currents is a consequence of the fact that the quarks' wavefunctions have all the same parity in $\Psi_{n\uparrow}^{0}$ and $\Psi_{p\uparrow}^{0}$. If they had different parity the combination would be axial-like charges with vector currents instead --eg., for a transition between  states with wavefunctions $\psi_{1,-1}^{s}$ and $\psi_{1,+1}^{s}$.

As for  the quadratic vacuum fluctuations of the boson fields which enter Eqs.(\ref{loparZ}), (\ref{loparW}), (\ref{loperZ}), (\ref{loperW}), writing them in terms of normal modes, and replacing the field Hamiltonian in those equations with their corresponding eigenvalues \cite{Peskin}, we may write 
\begin{equation}
	\left\langle V_{\mu}(\mathbf{x})\frac{1}{H_{F}^{V}}V_{\nu}^{\dagger}(\mathbf{y})\right\rangle=\int\frac{d^{3}q}{(2\pi)^{3}}\frac{e^{\pm i\mathbf{q}\cdot(\mathbf{x}-\mathbf{y})}}{-2(\mathcal{E}^{V}_{q})^{2}}\left(\eta_{\mu\nu}-\frac{q_{\mu}q_{\nu}}{M_{V}^{2}}\right),
\end{equation}
with  $V=\{Z,W^{\pm}\}$, and $\eta_{\mu\nu}$ being the Minkowski metric.

Lastly, inserting the values for $g_{v,a}^{u,d}$ \cite{Donoghe} in Eqs.(\ref{loparZ}), (\ref{loparW}), (\ref{loperZ}), (\ref{loperW})  we arrive at
\begin{eqnarray}
	\langle n^{0}_{\uparrow}|\mathbf{P}^{\textrm{Z}}_{\parallel}|n^{0}_{\uparrow}\rangle&=\frac{\hat{\mathbf{z}}5g^{2}}{24\cos^{2}{\theta_{W}}}\int\frac{d^{3}q}{(2\pi)^{3}}\int d^{3}x\int d^{3}y\:\frac{e^{i\mathbf{q}\cdot(\mathbf{x}-\mathbf{y})}\left(1+\frac{q_{z}^{2}}{M_{Z}^{2}}\right)}{2(q^{2}+M_{Z}^{2})}\nonumber\\
	&\times[F_{1-1}^{2}(x)+G_{1-1}^{2}(x)][F_{1-1}^{2}(y)+\cos{2\theta_{y}}G_{1-1}^{2}(y)],\label{loparZ2}\\
	\langle n^{0}_{\uparrow}|\mathbf{P}^{\textrm{W}}_{\parallel}|n^{0}_{\uparrow}\rangle&=\frac{\hat{\mathbf{z}}5g^{2}}{12}\int\frac{d^{3}q}{(2\pi)^{3}}\int d^{3}x\int d^{3}y\:\frac{e^{i\mathbf{q}\cdot(\mathbf{x}-\mathbf{y})}\left(1+\frac{q_{z}^{2}}{M_{W}^{2}}\right)}{2(q^{2}+M_{W}^{2})}\nonumber\\
	&\times[F_{1-1}^{2}(x)+G_{1-1}^{2}(x)][F_{1-1}^{2}(y)+\cos{2\theta_{y}}G_{1-1}^{2}(y)],\label{loparW2}
\end{eqnarray}
\begin{eqnarray}
	\langle n^{0}_{\uparrow}|\mathbf{P}^{\textrm{Z}}_{\perp}|n^{0}_{\uparrow}\rangle&=\frac{-\hat{\mathbf{z}}5g^{2}}{24\cos^{2}{\theta_{W}}}\int\frac{d^{3}q}{(2\pi)^{3}}\int d^{3}x\int d^{3}y\:\frac{q_{z}^{2}\:e^{i\mathbf{q}\cdot(\mathbf{x}-\mathbf{y})}}{2M_{Z}^{2}(q^{2}+M_{Z}^{2})}\nonumber\\
	&\times[F_{1-1}^{2}(x)+G_{1-1}^{2}(x)][F_{1-1}^{2}(y)+\cos{2\theta_{y}}G_{1-1}^{2}(y)],\label{loperZ2}\\
	\langle n^{0}_{\uparrow}|\mathbf{P}^{\textrm{W}}_{\perp}|n^{0}_{\uparrow}\rangle&=\frac{-\hat{\mathbf{z}}5g^{2}}{12}\int\frac{d^{3}q}{(2\pi)^{3}}\int d^{3}x\int d^{3}y\:\frac{q_{z}^{2}\:e^{i\mathbf{q}\cdot(\mathbf{x}-\mathbf{y})}}{2M_{W}^{2}(q^{2}+M_{W}^{2})}\nonumber\\
	&\times[F_{1-1}^{2}(x)+G_{1-1}^{2}(x)][F_{1-1}^{2}(y)+\cos{2\theta_{y}}G_{1-1}^{2}(y)],\label{loperW2}
\end{eqnarray}
for the longitudinal and transverse momenta, respectively. Comparing  Eqs.(\ref{loparZ2}) and (\ref{loparW2}) with Eqs.(\ref{loperZ2}) and (\ref{loperW2}) we note that the transverse momenta cancel out identically with the second $q$-dependent terms of the longitudinal momenta. Finally, we calculate the remaining integrals that account for the interaction of the transient chiral charges with the weak currents, mediated by Z and W-bosons. That is, 
\begin{eqnarray}
\fl	\int\frac{d^{3}q}{(2\pi)^{3}}\int d^{3}x\int d^{3}y\:\frac{e^{i\mathbf{q}\cdot(\mathbf{x}-\mathbf{y})}}{q^{2}+M_{V}^{2}}\mathcal{M}(x)\mathcal{N}(y)&=\int dx\:x^{2}\int dy\:y^{2}\frac{2\pi}{M_{V}xy}\\
\fl	&\times\left[
	e^{-M_{V}|x-y|}-e^{-M_{V}(x+y)}\right]\mathcal{M}(x)\mathcal{N}(y)\nonumber\\
\fl	&\simeq\frac{4\pi}{M_{V}}\int dx\:x^{2}\mathcal{M}(x)\mathcal{N}(x),\quad V=\{Z,W^{\pm}\},\nonumber
\end{eqnarray}
where in the last approximation we have made use of the fact that the integrals extend over a distance of the order of the neutron radius, and $R_{n}\gg1/M_{Z,W}$. The insertion of this formula into Eqs.(\ref{loparZ2}), (\ref{loparW2}), (\ref{loperZ2}) and (\ref{loperW2}) leads to the expression for $\mathcal{F}_{n}$ of Eq.(\ref{result1}) in Section \ref{EWmomentum}. The numerical estimations were made there, for the MIT-bag model, with $F_{1-1}(x)=j_{0}(2.04\:x/R_{n}), G_{1-1}(x)=j_{1}(2.04\:x/R_{n})$  and $R_{n}=1.1$ fm.

\section{Nuclear Force}\label{appforce}
It has been claimed in Section \ref{EWmomentum} that, throughout the spin reversal process of the neutron with the RF field, the nuclear force along $\hat{\bf{z}}$ is $\omega_{1}\sin{(\omega_{1}t)}\Delta\langle P^{\textrm{sf}}_{\textrm{kin}}\rangle/2$, with  $\omega_{1}=\gamma_{n}B_{1}$. It has been also stated in Section \ref{MomentumK} that this force  possesses a conservative component  associated to the variation of the transverse momenta of the boson fields, and a non-conservative component related to the variation of the longitudinal momenta. Here we outline the proof of these assertions.

Let us start with Eq.(\ref{primera}), 
\begin{eqnarray}
	\fl	\langle F_{z}(t)\rangle=\frac{d}{dt}\langle P^{\textrm{kin}}_{z}(t)\rangle&=-\frac{d}{dt}\left[\langle P^{\textrm{Z}}_{\parallel,z}(t)\rangle+\langle P^{\textrm{W}}_{\parallel,z}(t)\rangle+\langle P^{\textrm{Z}}_{\perp,z}(t)\rangle+\langle P^{\textrm{W}}_{\perp,z}(t)\rangle\right]\nonumber\\
\fl	&\equiv\langle F_{\parallel,z}(t)\rangle+\langle F_{\perp,z}(t)\rangle,\nonumber
\end{eqnarray}
where, after the last equality, we have identified the force components associated to the time variation of the longitudinal and the transverse momenta of the gauge bosons. In the following, we restrict ourselves to the contribution of the momenta of $Z$ bosons. The computation is in all analogous for that of $W$ bosons. Starting with the expression of Eq.(\ref{Ppar}) for $	\langle P^{\textrm{Z}}_{\parallel,z}(t)\rangle$, and restricting the calculation to the $z$ component of the force, we write
\begin{eqnarray}
\fl	\langle P^{\textrm{Z}}_{\parallel,z}(t)\rangle&=
	2\textrm{Re}\langle n^{0}_{\downarrow}|\cos{(\omega_{1}t/2)}e^{im_{n}t}P^{\textrm{Z}}_{\parallel,z}
	\int _{0}^{t}-i\:d\tau e^{-\Gamma_{Z}(t-\tau)/2}\mathbb{U}_{0}(t-\tau)W_{nc}\cos{(\omega_{1}\tau/2)}\nonumber\\
	\fl &\times e^{-im_{n}\tau}|n^{0}_{\downarrow}\rangle
	+2\textrm{Re}\langle n^{0}_{\uparrow}|\sin{(\omega_{1}t/2)}e^{im_{n}t}P^{\textrm{Z}}_{\parallel,z}
	\int _{0}^{t}-i\:d\tau e^{-\Gamma_{Z}(t-\tau)/2}\mathbb{U}_{0}(t-\tau)\nonumber\\
	\fl&\times W_{nc}\sin{(\omega_{1}\tau/2)}e^{-im_{n}\tau}|n^{0}_{\uparrow}\rangle.\label{PparZ} 
\end{eqnarray}
In this equations, using Eq.(\ref{nt}), we have replaced the time-propagators $\mathbb{U}_{0}^{\dagger}(t)$ and $\mathbb{U}_{0}(\tau)$ in Eq.(\ref{Ppar}) with their explicit expressions as a function of time. Note also that cross terms of the form $\langle n^{0}_{\downarrow}|...\mathbf{P}^{\textrm{Z}}_{\parallel}...|n^{0}_{\uparrow}\rangle$ do not contribute to $\langle P^{\textrm{Z}}_{\parallel,z}(t)\rangle$. Straightforward integration of this expression for $t\gg\Gamma_{Z}^{-1}$ yields $\langle P^{\textrm{Z}}_{\parallel,z}(t)\rangle=\langle P^{\textrm{Z},\downarrow}_{\parallel,z}\rangle\cos{(\omega_{1}t)}$, where $\langle P^{\textrm{Z},\downarrow}_{\parallel,z}\rangle$ is the longitudinal momentum for the spin-down configuration.   Next, for the computation of 
$\langle F_{\parallel,z}(t)\rangle=-(d/dt)(\langle P^{\textrm{Z}}_{\parallel,z}(t)\rangle+\langle P^{\textrm{W}}_{\parallel,z}(t)\rangle)$, we distinguish three kinds of  time-derivative terms from Eq.(\ref{PparZ}), namely, (1) the one that involves  the time integrals in $\tau$, (2) the one which applies to $\cos{(\omega_{1}t/2)},\sin{(\omega_{1}t/2)}$ which accompany the neutron states on the left, and (3) the one which applies to $e^{-\Gamma_{Z}(t-\tau)/2}\mathbb{U}_{0}(t-\tau)$ within the integrands. The last one, (3), cancels out with one of the terms in (2); those terms of kind (2) which are  proportional to $\Gamma_{Z}\langle P^{\textrm{Z},\downarrow}_{\parallel,z}\rangle$ cancel out; and the remainder of terms of kinds (1) and (3) amounts to 
\begin{equation}
	\frac{d}{dt}\langle P^{\textrm{Z}}_{\parallel,z}(t)\rangle=-\omega_{1}\langle P^{\textrm{Z},\downarrow}_{\parallel,z}\rangle\sin{(\omega_{1}t)},\label{segunda}
\end{equation}
which is straightforward identifiable with $-2\langle P^{\textrm{Z}\downarrow}_{\parallel,z}\rangle\frac{d}{dt}\langle S_{n,z}(t)\rangle$. This is a  nonconservative force.

In regards to the transverse components of the force, we write analogously,
\begin{eqnarray}
\fl	\langle P^{\textrm{Z}}_{\perp,z}(t)\rangle&=\langle n^{0}_{\downarrow}|\int _{0}^{t}d\tau' 
	\cos{(\omega_{1}\tau'/2)}e^{im_{n}\tau'}W^{\dagger}_{nc}e^{-\Gamma_{Z}(t-\tau')/2}
	\mathbb{U}_{0}^{\dagger}(t-\tau')\:P^{\textrm{Z}}_{\perp,z}\nonumber\\
\fl	&\times\int _{0}^{t}d\tau e^{-\Gamma_{Z}(t-\tau)/2}\mathbb{U}_{0}(t-\tau)W_{nc}\mathbb{U}_{0}(\tau)\cos{(\omega_{1}\tau/2)}|n^{0}_{\downarrow}\rangle\nonumber\\
\fl	&+\langle n^{0}_{\uparrow}|\int _{0}^{t}d\tau' 
	\sin{(\omega_{1}\tau'/2)}\mathbb{U}_{0}^{\dagger}(\tau' )W^{\dagger}_{nc}e^{-\Gamma_{Z}(t-\tau')/2}
	\mathbb{U}_{0}^{\dagger}(t-\tau')\:P^{\textrm{Z}}_{\perp,z}\nonumber\\
\fl	&\times\int _{0}^{t}d\tau e^{-\Gamma_{Z}(t-\tau)/2}\mathbb{U}_{0}(t-\tau)W_{nc}e^{-im_{n}\tau}\sin{(\omega_{1}\tau/2)}|n^{0}_{\uparrow}\rangle.\label{PperZ}
\end{eqnarray} 
Again, straightforward integration of this expression for $t\gg\Gamma_{Z}^{-1}$ yields $\langle P^{\textrm{Z}}_{\perp,z}(t)\rangle=\langle P^{\textrm{Z},\downarrow}_{\perp,z}\rangle\cos{(\omega_{1}t)}$, where $\langle P^{\textrm{Z},\downarrow}_{\perp,z}\rangle$ is the transverse momentum for the spin-down configuration. Once more, for the computation of 
$\langle F_{\perp,z}(t)\rangle=-(d/dt)(\langle P^{\textrm{Z}}_{\perp,z}(t)\rangle+\langle P^{\textrm{W}}_{\perp,z}(t)\rangle)$, we distinguish two kinds of  time-derivative terms from Eq.(\ref{PperZ}), namely, (1) the one that involves  the time integrals in $\tau^{(')}$, and (2) the one which applies to $e^{-\Gamma_{Z}(t-\tau^{(')})/2}\mathbb{U}_{0}(t-\tau^{(')})$ within the integrands. As for the first kind, (1), following steps analogous to those in Ref.\cite{Donaire_vdW}, they are easily identifiable with 
$\langle \nabla_{z}W_{nc}(t)\rangle$, where the subscript $z$ refers to the $z$-component of the center of mass position vector \footnote{This assertion, however, relies on the possibility to separate the Hamiltonian in external and internal degrees of freedom, which is the case of the HO model for quark confinement.}. This, however, contains both a term which equals $-\omega_{1}\langle P^{\textrm{Z},\downarrow}_{\perp,z}\rangle\sin{(\omega_{1}t)}$, and another one of the form $\Gamma_{Z}\langle P^{\textrm{Z},\downarrow}_{\perp,z}\rangle\cos{(\omega_{1}t)}$ due to the presence of the unstable $Z$ particle in the final state of the quantum process. The latter  cancels with the contribution of the terms  of kind (2), of equal strength and opposite sign. All in all the end result is 
\begin{equation}
	\frac{d}{dt}\langle P^{\textrm{Z}}_{\perp,z}(t)\rangle=-\omega_{1}\langle P^{\textrm{Z},\downarrow}_{\perp,z}\rangle\sin{(\omega_{1}t)},\label{tercera}
\end{equation}
which is again  identifiable with $-2\langle P^{\textrm{Z}\downarrow}_{\perp,z}\rangle\frac{d}{dt}\langle S_{n,z}(t)\rangle$. Since it derives from those terms which form part of $\langle \nabla_{z}W_{nc}(t)\rangle$, we deduce that it is a conservative force, as stated in the text.

Finally, putting together Eqs.(\ref{primera}), (\ref{segunda}) and (\ref{tercera}), we end up with the result of Section \ref{EWmomentum},
\begin{small}
\begin{equation}
\fl	\langle F_{z}(t)\rangle=\omega_{1}\left[\langle P^{\textrm{Z},\downarrow}_{\parallel,z}\rangle+\langle P^{\textrm{W},\downarrow}_{\parallel,z}\rangle+\langle P^{\textrm{Z},\downarrow}_{\perp,z}\rangle+\langle P^{\textrm{W},\downarrow}_{\perp,z}\rangle\right]\sin{(\omega_{1}t)}=\omega_{1}\sin{(\omega_{1}t)}\Delta\langle P^{\textrm{sf}}_{\textrm{kin}}\rangle/2.
\end{equation}
\end{small}
\section{Energetics}
It has been stated in the Conclusions Section that a calculation analogous to that of Ref.\cite{JPCMDonaire} for the charges bound in a magnetochiral molecule but for the constituent quarks of a neutron  reveals that the kinetic energy gained by the neutron corresponds to corrections to its EW self-energy, i.e., corrections to the energy associated to the EW interactions between the constituent quarks of the neutron, $\langle n^{0}(t)|W_{EM}+W_{nc}+W_{cc}|n^{0}(t)\rangle$. Also, it is claimed that the source of this energy is the external magnetic field which causes its spin reversal. In the following, we explain these statements.

\subsection{Kinetic energy as a correction to the EW self-energy}
The corrections to the EW self-energy  of the neutron we are interested in are those induced by the motion of its centre of mass. 

In the first place, using the nonrelativistic harmonic-oscillator model and Jacobi's coordinates outlined in \ref{gauss}, one finds that the interaction Hamiltonian densities $\mathcal{W}_{EM}$,   $\mathcal{W}_{cc}$ and  $\mathcal{W}_{nc}$ contain terms of the form $-\mathbf{P}\cdot\boldsymbol{\mathcal{P}}^{A}_{\parallel}/m_{n}$, $-\mathbf{P}\cdot\boldsymbol{\mathcal{P}}^{W}_{\parallel}/m_{n}$ and  $-\mathbf{P}\cdot\boldsymbol{\mathcal{P}}^{Z}_{\parallel}/m_{n}$, respectively, which couple the canonical conjugate momentum of the neutron centre of mass to the longitudinal momenta of each boson field. A straightforward calculation of the variation of the expectation values of these terms along the spin reversal process, which is analogous to that in Ref.\cite{JPCMDonaire} for the EM self-energy of a magnetochiral molecule along the adiabatic switching of an external magnetic field, yields, at leading order in the coupling constants, 
\begin{equation}
\fl	\delta\mathcal{E}_{\parallel}(t)=[\mathbf{P}_{\textrm{kin}}^{0}-\Delta\langle\mathbf{P}_{\perp}^{W}+\mathbf{P}_{\perp}^{Z}+\mathbf{P}_{\parallel}^{W}+\mathbf{P}_{\parallel}^{Z}\rangle(t)]\cdot[\delta\langle\mathbf{P}_{\parallel}^{W}(t)\rangle+\delta\langle\mathbf{P}_{\parallel}^{Z}(t)\rangle]/m_{n}.\label{lambparal}
\end{equation}
The reason for referring to this quantity as EW self-energy with subscript $\parallel$, $\delta\mathcal{E}_{\parallel}$, is that it results from the internal EW interactions between the constituent quarks and derives from the coupling of the total momentum of the neutron to the longitudinal momenta of the EW boson fields. Note that in our case $\langle\mathbf{P}_{\parallel}^{A}\rangle=\mathbf{0}$ at any time, which implies that the EM self-energy can be neglected and we can restrict ourselves to the computation of the weak nuclear self-energy, i.e., that which accounts for the chiral interactions only. On the right hand side of Eq.(\ref{lambparal}), in the first factor within square brackets, $\mathbf{P}_{\textrm{kin}}^{0}$ is the initial kinetic momentum of the neutron, and Eq.(\ref{Pconv}) of Section \ref{MomentumK} for total momentum conservation has been applied. In the second factor, the variations of the longitudinal momenta read $\delta\langle\mathbf{P}_{\parallel}^{W,Z}(t)\rangle=(\partial\langle\mathbf{P}_{\parallel}^{W,Z}(t)\rangle/\partial t)\delta t$, which are proportional to the time variation of the neutron spin expectation value, $\delta\langle\mathbf{S}_{n}(t)\rangle$. Hence direct integration of this expression along the spin flip (sf) process yields
\begin{equation}
\fl	\Delta^{\textrm{sf}}\mathcal{E}_{\parallel}=-[\mathbf{P}_{\textrm{kin}}^{0}-\Delta^{\textrm{sf}}\langle\mathbf{P}^{W}_{\perp}+\mathbf{P}^{Z}_{\perp}+\mathbf{P}_{\parallel}^{W}+\mathbf{P}_{\parallel}^{Z}\rangle/2]\cdot\Delta^{\textrm{sf}}\langle\mathbf{P}_{\parallel}^{W}+\mathbf{P}_{\parallel}^{Z}\rangle/m_{n}.\label{Eparal}
\end{equation}

On the other hand, the frequency of the virtual bosons with any momentum $\mathbf{q}$ which mediate the EW interactions between the constituent quarks at second order in $\mathcal{W}_{EM},\mathcal{W}_{nc},\mathcal{W}_{cc}$, gets Doppler-shifted as a result of the non-null velocity of the centre of mass of the neutron. In turn, it gives rise to  finite corrections to the EW self-interaction mediated by loops of those bosons, and is proportional to their transverse momenta. The latter has been proved not to be null for the Z and W virtual bosons for the case of a polarized neutron. Hence, hereafter it suffices to consider the weak nuclear self-energy and to neglect the EM one. The expression of the weak nuclear self-energy of a polarized neutron at a certain time $t$, at second order in $\mathcal{W}_{nc},\mathcal{W}_{cc}$, is
\begin{eqnarray}
\fl	\mathcal{E}(t)&=\langle n^{0}(t)|W_{nc}+W_{cc}|n^{0}(t)\rangle=\langle n^{0}(t)|W^{\dagger}_{nc}\frac{1}{E_{0}(t)-H^{HO}-H_{F}^{Z}}W_{nc}|n^{0}(t)\rangle\nonumber\\
\fl	&+\langle n^{0}(t)|W^{\dagger}_{cc}\frac{1}{E_{0}(t)-H^{HO}-H_{F}^{W}}W_{cc}|n^{0}(t)\rangle,
\end{eqnarray}
where $E_{0}(t)$ is the total energy of the neutron at time $t$, $E_{0}(t)=m_{n}+\langle\mathbf{P}_{\textrm{kin}}(t)\rangle^{2}/2m_{n}$. 
In this equation the wavefunction of the free neutron states incorporate the prefactor $e^{-i\langle\mathbf{P}_{\textrm{kin}}(t)\rangle\cdot\mathbf{R}}/\sqrt{\mathcal{V}}$ of Eq.(\ref{wavef}) and $H^{HO}$ in the denominator contains the kinetic energy of its center of mass. Using the completeness relationship for the intermediate states of nucleons of internal level $m$ and spin $s$ (proton $p$, neutron $n$) and boson fields of momentum $\mathbf{q}$ and polarization $\lambda$, $\{|n_{m},s;\mathbf{q},\lambda\rangle,|p_{m},s;\mathbf{q},\lambda\rangle\}$, with energies $\mathcal{E}^{p,n}_{m}+\mathcal{E}^{W,Z}_{q}$ respectively, and applying conservation of total momentum, the above equation can be rewritten as--see Refs.\cite{JHEP,JPCMDonaire},
\begin{eqnarray}
	\mathcal{E}(t)&=\sum_{m,s,\mathbf{q},\lambda}\frac{\langle n^{0}(t)|W^{\dagger}_{nc}|n_{m},s;\mathbf{q},\lambda\rangle \langle n_{m},s;\mathbf{q},\lambda|W_{nc}|n^{0}(t)\rangle}{E_{0}(t)-\mathcal{E}^{n}_{m}-\mathcal{E}^{Z}_{q}+\mathbf{q}\cdot\langle\mathbf{P}_{\textrm{kin}}(t)\rangle/m_{n}}\nonumber\\
	&+\frac{\langle n^{0}(t)|W^{\dagger}_{cc}|p_{m},s;\mathbf{q},\lambda\rangle \langle p_{m},s;\mathbf{q},\lambda|W_{cc}|n^{0}(t)\rangle}{E_{0}(t)-\mathcal{E}^{p}_{m}-\mathcal{E}^{W}_{q}+\mathbf{q}\cdot\langle\mathbf{P}_{\textrm{kin}}(t)\rangle/m_{n}},
\end{eqnarray}
where the terms $\mathbf{q}\cdot\langle\mathbf{P}_{\textrm{kin}}(t)\rangle/m_{n}$ in the denominators are the energies associated to the Doppler shift induced by the motion of the neutron on the frequencies of the virtual bosons which mediate the internal interactions between its constituent quarks. Here, at any given time $t$,  $\langle\mathbf{P}_{\textrm{kin}}(t)\rangle$ is computed integrating Eq.(\ref{primera}). Next, treating the Doppler shift terms as small perturbations upon the leading terms $\mathcal{E}^{W,Z}_{q}$ and considering infinitesimal variations on the neutron states, we end up with
\begin{eqnarray}
\fl	\delta\mathcal{E}_{\perp}(t)&=-\frac{\langle\mathbf{P}_{\textrm{kin}}(t)\rangle}{m_{n}}\cdot\Bigl[\sum_{m,s,\mathbf{q},\lambda}2\textrm{Re}\frac{\delta\langle n^{0}(t)|W^{\dagger}_{nc}|n_{m},s;\mathbf{q},\lambda\rangle\mathbf{q}\langle n_{m},s;\mathbf{q},\lambda|W_{nc}|n^{0}(t)\rangle}{[E_{0}(t)-\mathcal{E}^{n}_{m}-\mathcal{E}^{Z}_{q}]^{2}}\nonumber\\
\fl	&+\sum_{m,s,\mathbf{q},\lambda}2\textrm{Re}\frac{\delta\langle n^{0}(t)|W^{\dagger}_{cc}|p_{m},s;\mathbf{q},\lambda\rangle\mathbf{q}\langle p_{m},s;\mathbf{q},\lambda|W_{cc}|n^{0}(t)\rangle}{[E_{0}(t)-\mathcal{E}^{p}_{m}-\mathcal{E}^{W}_{q}]^{2}}\Bigr],\label{lambperp}
\end{eqnarray}
where $\delta\langle n^{0}(t)|$ stands for an infinitesimal variation on the neutron state caused by the coherent rotation of its spin under the action of the external magnetic fields. Direct comparison of Eq.(\ref{lambperp}) with Eqs.(\ref{Ppar2a}) and (\ref{Ppar2b}) allows us to identify the terms under the summation symbols with $\delta\langle\mathbf{P}_{\perp}^{W}\rangle+\delta\langle\mathbf{P}_{\perp}^{Z}\rangle$. This explains the subscript $\perp$ in $\delta\mathcal{E}_{\perp}(t)$, as it is proportional to the variation of the transverse momenta. Direct integration of Eq.(\ref{lambperp}) along the spin flip (sf) process yields
\begin{equation}
\fl	\Delta^{\textrm{sf}}\mathcal{E}_{\perp}=-[\mathbf{P}_{\textrm{kin}}^{0}-\Delta^{\textrm{sf}}\langle\mathbf{P}^{W}_{\perp}+\mathbf{P}^{Z}_{\perp}+\mathbf{P}_{\parallel}^{W}+\mathbf{P}_{\parallel}^{Z}\rangle/2]\cdot\Delta^{\textrm{sf}}\langle\mathbf{P}_{\perp}^{W}+\mathbf{P}_{\perp}^{Z}\rangle/m_{n},\label{Eperp}
\end{equation}
where $\mathbf{P}_{\textrm{kin}}^{0}$ is the initial kinetic momentum of the neutron and total momentum conservation has been applied again.

Finally, putting Eqs.(\ref{Eparal}) and (\ref{Eperp}) together we end up with
\begin{eqnarray}
	\delta\mathcal{E}_{\parallel}(t)+\delta\mathcal{E}_{\perp}(t)&=-\mathbf{P}_{\textrm{kin}}^{0}\cdot\Delta^{\textrm{sf}}\langle\mathbf{P}_{\parallel}^{W}+\mathbf{P}_{\parallel}^{Z}+\mathbf{P}_{\perp}^{W}+\mathbf{P}_{\perp}^{Z}\rangle/m_{n}\nonumber\\
	&+\left[\Delta^{\textrm{sf}}\langle\mathbf{P}_{\parallel}^{W}+\mathbf{P}_{\parallel}^{Z}+\mathbf{P}_{\perp}^{W}+\mathbf{P}_{\perp}^{Z}\rangle\right]^{2}/2m_{n},\label{Lamb}
\end{eqnarray}
which equals the kinetic energy gained by the neutron along the reversal of its spin, as claimed in the Conclusions. 

\subsection{On the source of the kinetic energy and its adiabatic variations}\label{energetics_source}

As for the source of the kinetic energy, it can be inferred from Eqs.(\ref{lambparal}) and (\ref{lambperp}) and the fact that  
\begin{equation}
	\delta\langle\mathbf{P}_{\parallel,\perp}^{W,Z}(t)\rangle=(\partial\langle\mathbf{P}_{\parallel,\perp}^{W,Z}(t)\rangle/\partial t)\delta t\propto\delta\langle\mathbf{S}_{n}(t)\rangle,
\end{equation}
that the kinetic energy is provided by the sources of the external magnetic field that causes the spin rotation. Hence, the energy in Eq.(\ref{Lamb}) can be also interpreted as a magnetisation energy, as it contains terms that go like  $\sim\mathbf{P}_{\textrm{kin}}^{0}\cdot\Delta^{\textrm{sf}}\langle\mathbf{S}_{n}\rangle$ and $\sim\Delta^{\textrm{sf}}\langle\mathbf{S}_{n}\rangle\cdot\Delta^{\textrm{sf}}\langle\mathbf{S}_{n}\rangle$. In contrast to the kinetic energy gained by a magnetochiral molecule or by a non-polarized proton \cite{JPCMDonaire,JHEP}, where the magnetization is caused by an external magnetic field which is adiabatically switched on, it is caused here by the combination of the uniform field $\mathbf{B}_{0}$ and the RF field $\mathbf{B}_{1}(t)$ whose time variation is adiabatic with respect to the dynamics of the internal degrees of freedom of the system neutron-gauge fields. 

On the other hand, it is the adiabatic rotation of the spin that guarantees the perturbative computation of $\delta\langle\mathbf{P}_{\parallel,\perp}^{W,Z}(t)\rangle$ and $\delta\mathcal{E}_{\parallel,\perp}(t)$. That is, in the computation of all these quantities, the lifetime of the intermediate states is $\sim\Gamma_{Z,W}^{-1}$, which is much less than the time scale of the spin variations, $\Gamma_{Z,W}^{-1}\ll\omega_{1}^{-1}$. Therefore, at any given time $t$, the variations of momentum and energy take place in a time interval  $\delta t\sim\Gamma_{Z,W}^{-1}$. Since the nuclear forces responsible for these variations scale as   $\omega_{1}\langle\mathbf{P}_{\parallel,\perp}^{\downarrow,W,Z}\rangle$ --see \ref{appforce}, the variations of momenta and energy within each time-interval $\delta t$ are of the order of $(\omega_{1}/\Gamma_{Z,W})\Delta^{\textrm{sf}}\langle\mathbf{P}_{\parallel,\perp}^{W,Z}\rangle$ and $(\omega_{1}/\Gamma_{Z,W})[\mathbf{P}_{\textrm{kin}}^{0}\cdot\Delta^{\textrm{sf}}\langle\mathbf{P}_{\parallel,\perp}^{W,Z}\rangle/m_{n}+(\Delta^{\textrm{sf}}\langle\mathbf{P}_{\parallel,\perp}^{W,Z}\rangle)^{2}/2m_{n}]$, respectively. In particular, the variation of the energy is much less than the magnetic energy associate to the spin flip, $\omega_{0}$.

\end{document}